\documentclass[usenatbib,usegraphicx]{mn2e}
\usepackage{amssymb}
\usepackage{graphicx}
\usepackage{mathrsfs}
\usepackage{longtable}

\newcommand{\OVIdblt}{{O}\kern 0.1em{\sc vi}~$\lambda\lambda 1032, 1038$}
\newcommand{\CII}{\hbox{{C}\kern 0.1em{\sc ii}}}
\newcommand{\CIII}{\hbox{{C}\kern 0.1em{\sc iii}}}
\newcommand{\CIV}{\hbox{{C}\kern 0.1em{\sc iv}}}
\newcommand{\HI}{\hbox{{H}\kern 0.1em{\sc i}}}
\newcommand{\Lya}{\hbox{{Ly}\kern 0.1em$\alpha$}}
\newcommand{\Lyb}{\hbox{{Ly}\kern 0.1em$\beta$}}
\newcommand{\Lyg}{\hbox{{Ly}\kern 0.1em$\gamma$}}
\newcommand{\Lyd}{\hbox{{Ly}\kern 0.1em$\delta$}}
\newcommand{\Lye}{\hbox{{Ly}\kern 0.1em$\epsilon$}}
\newcommand{\Lyz}{\hbox{{Ly}\kern 0.1em$\zeta$}}
\newcommand{\Lyeta}{\hbox{{Ly}\kern 0.1em$\eta$}}
\newcommand{\MgII}{\hbox{{Mg}\kern 0.1em{\sc ii}}}
\newcommand{\OVI}{\hbox{{O}\kern 0.1em{\sc vi}}}
\newcommand{\OVII}{\hbox{{O}\kern 0.1em{\sc vii}}}
\newcommand{\OVIII}{\hbox{{O}\kern 0.1em{\sc viii}}}
\newcommand{\NV}{\hbox{{N}\kern 0.1em{\sc v}}}
\newcommand{\SiII}{\hbox{{Si}\kern 0.1em{\sc ii}}}
\newcommand{\SiIII}{\hbox{{Si}\kern 0.1em{\sc iii}}}
\newcommand{\SiIV}{\hbox{{Si}\kern 0.1em{\sc iv}}}
\newcommand{\FeII}{\hbox{{Fe}\kern 0.1em{\sc ii}}}

\newcommand{\NeIX}{\hbox{{Ne}\kern 0.1em{\sc ix}}}
\newcommand{\OI}{\hbox{{O}\kern 0.1em{\sc i}}}

\title[Impact of UCMHs on the IGM and First Structure]
{Impact of Primordial Ultracompact Minihaloes on the Intergalactic Medium and First Structure Formation}
\author[Dong Zhang]{Dong Zhang$^1$\thanks{dzhang@astronomy.ohio-state.edu}\\
$^1$Department of Astronomy, Ohio State University, 140 W. 18th Ave., Columbus, OH 43210, USA}

\begin{document}

\maketitle

\begin{abstract}

The effects of dark matter annihilation on the evolution of
intergalactic medium (IGM) in the early Universe can be more
important if dark matter structure is more concentrated.
Ultracompact Minihaloes (UCMHs), which formed by dark matter
accretion onto primordial black holes (PBHs) or initial dark matter
overdensity produced by the primordial density perturbation, provide
a new type of compact dark matter structure to ionize and heat the
IGM after matter-radiation equality $z_{\rm eq}$, which is much
earlier than the formation of the first cosmological dark halo
structure and later first stars. We show that dark matter
annihilation density contributed by UCMHs can totally dominated over
the homogenous dark matter annihilation background even for a tiny UCMH
fraction $f_{\rm UCMH}=\Omega_{\rm UCMH}(z_{\rm eq})/\Omega_{\rm
DM}\geq10^{-15}(1+z)^{2}(m_{\chi}c^{2}/100$ GeV$)^{-2/3}$ with a
standard thermal relic dark matter annihilation cross section, and provide a new
gamma-ray background in the early Universe. UCMH annihilation becomes important
to the IGM evolution approximately for $f_{\rm UCMH}>10^{-6}(m_{\chi}c^{2}/100$ GeV).
The IGM ionization fraction $x_{\rm ion}$ and gas temperature $T_{\rm m}$ can be
increased from the recombination residual $x_{\rm ion}\sim10^{-4}$ and adiabatically
cooling $T_{\rm m}\propto(1+z)^{2}$ in the absence of energy injection, to the
highest value of $x_{\rm ion}\sim0.1$ and $T_{\rm m}\sim 5000$ K at $z\geq10$
for the upper bound UCMH abundance constrained by the cosmic microwave background
optical depth.

A small fraction of UCMHs are seeded by PBHs. The X-ray emission
from gas accretion onto PBHs may totally dominated over
dark matter annihilation and become the main cosmic ionization source for
a PBH abundance $f_{\rm PBH}=\Omega_{\rm PBH}/\Omega_{\rm DM}\gg10^{-11}$ $(10^{-12})$ with
the PBH mass $M_{\rm PBH}\sim10^{-6}M_{\odot}$ $(10^{2}M_{\odot})$.
However, the constraints of gas accretion rate and X-ray absorption
by the baryon accumulation within the UCMHs and accretion feedback show that X-ray emission
can only be a promising source much later than UCMH annihilation at
$z<z_{m}\ll1000$, where $z_{m}$ depends on the masses of PBHs, their host UCMHs,
and the dark matter particles. Also, UCMH radiation including both annihilation
and X-ray emission can significantly suppress the low mass first baryonic structure formation.
The effects of UCMHs radiation on the baryonic structure evolution are
quite small for the gas temperature after virialization, but more significant to enhance the gas
chemical quantities such as the ionization fraction and molecular hydrogen abundance
in the baryonic objects.

\end{abstract}

\begin{keywords}
intergalactic medium --- dark matter --- cosmology: theory
--- early Universe --- galaxies: structure
\end{keywords}

\section{Introduction}\label{intro}

Ultracompact Minihaloes (UCMHs) are primordial dark matter
structures which formed by dark matter accreting onto primordial
black holes (PBHs) after matter-radiation equality $z_{\rm
eq}\sim3100$, or direct collapsed onto an initial dark matter
overdensity produced by small density perturbation before $z_{\rm
eq}$, e.g., in several Universe phase transition epochs
(\citealt{Mack07}; \citealt{Ricotti09}). If the density perturbation
in the early Universe exceed a critical value
$\delta_{c}=(\delta\rho/\rho)_{c}\sim1/3$, this region becomes
gravitationally unstable and directly collapse to form a PBH
(\citealt{Hawking71}; \citealt{Carr74}; see \citealt{Khlopov10} for
a review and references therein). PBHs which form with a sufficient
high mass $\geq10^{16}$ gram do not evaporate but begin to grow by
accreting the surrounding dark matter and form a compact dark matter
halo, which will grow by two orders of magnitude in mass
during the matter dominated era (\citealt{Mack07}). These haloes are so-called Ultracompact
Minihaloes (UCMHs), or say Primordially-Laid Ultracompact Minihaloes
(PLUMs). On the other hand, small density perturbation in the early
Universe $10^{-3}<\delta<\delta_{c}$ will form a compact dark matter
overdensity instead of a PBH. Such an overdense cloud can also seed
the formation of UCMHs (\citealt{Ricotti09}; \citealt{Scott09};
\citealt{Josan10}). Note that the initial density perturbations from
inflation were just $\delta\sim10^{-4}-10^{-5}$, it is proposed
UCMHs are far more viable to form by accreting onto dark matter
overdensity, which requires a much lower perturbation threshold than
PBHs. Also, the UCMHs seeded by primordial overdensities have a
different profile with those seeded by PBHs (\citealt{Bert85};
\citealt{Mack07}).

UCMHs have been recently proposed as a new type of non-baryonic
massive compact gravitational object (MACHO; \citealt{Ricotti09}) as
well as gamma-ray and neutrino source (\citealt{Scott09}). UCMHs
could produce a microlensing lightcurve which can be distinguished
from that of a ``point-like'' object such as a star or brown dwarf,
thus become a promising new target for microlensing searches.
Moreover, the abundance of UCMHs can be constrained by the
observation of the Milky Way gamma-ray flux and the extragalactic
gamma-ray background, although this constraint is still very uncertain based on
today's data (\citealt{Lacki10}; \citealt{Josan10};
\citealt{Saito11}). Since we know the growth of an isolate UCMH as a
function of redshift (\citealt{Mack07}), we can natively trace the
fraction of UCMHs back to very high redshift without considering
mergers and tidal destruction. Until now, most of the works on UCMHs
focus on the properties of the nearby UCMHs at $z<1$. Another important question which has barely been
discussed is that, what are the consequences of UCMH radiation at
very high redshift, since UCMHs are the ``remnants'' originally from
the early Universe? As the sources of heating and ionization
before the first structure, stars and galaxies, sufficient
UCMHs may play important roles to change the chemical and thermal
history of the early Universe. Our main purpose in this paper is to
investigate the impacts of UCMH emission on the intergalactic medium
(IGM) in the Universe reionization era, and the following first baryon structure
formation and evolution.

The process of reionization of all hydrogen atoms in the IGM
would have been completed at redshift $z\approx6$ (e.g.,
\citealt{Becker01}; \citealt{Fan02}). However, much earlier
ionization at $z>6$ is implied by the WMAP observation (e.g.,
\citealt{Dunkley09}; \citealt{Komatsu09}). It is commonly suggested
that the possible contributions to the high redshift reionization
between approximately $6<z<20$ are the first baryonic objects to
produce significant ultraviolet light, early (Pop III and Pop II)
stars, and old quasars (\citealt{BL07}; \citealt{Wise08};
\citealt{VG09}; \citealt{Meiksin09}). However, it is still unclear
whether quasars and first stars were sufficiently efficiency to
reionize the universe. Dark matter, on the other hand, is suggested
to be the exotic source of ionization and heating
at high redshift due to its self-annihilation or decay. It is
usually proposed that weakly interacting massive particles (WIMPs)
provide a compelling solution to identify the dark matter component.
The mass of the dark matter particles $m_{\chi}$, and the average
annihilation cross section $\langle\sigma v\rangle$ are the two
crucial parameters to affect the ionizing and heating processes.
Under the thermal relic assumption that the cross section
$\langle\sigma v\rangle\approx3\times10^{-26}$ cm$^{3}$ s$^{-1}$ to
match the observed $\Omega_{\rm DM}h^{2}\approx0.110$, most previous
studies showed that the effects of homogenous dark matter background
annihilation or decay on the high-redshift IGM are expected to be
important only for light dark matter $m_{\chi}c^{2}\leq1$ GeV or
sterile neutrinos (e.g., \citealt{Hansen04}; \citealt{Pierpaoli04};
\citealt{Mapelli05}; \citealt{Belotsky05}; \citealt{Zhang06};
\citealt{Mapelli06}; \citealt{RMF07a, RMF07b}; \citealt{Chluba10}).
The annihilation flux would be enhanced only after the formation of
the first dark objects $z<60$, as dark matter become more clumpy
(e.g., \citealt{Chuzhoy08}; \citealt{NS08,NS09,NS10}; \citealt{BH09,
BH10}). However, in our case, as dark matter is more concentrated in
UCMHs which are significantly denser than the homogenous dark matter
background, WIMP dark matter annihilation within UCMHs may become
powerful gamma-ray source dominated over the homogenous background
annihilation, even though UCMHs are very rare.

A small fraction of UCMHs are seeded by PBHs (\citealt{Mack07}). In this paper we call
these UCMHs as PBH host UCMHs. Since PBH abundance is still uncertain for a broad
range of PBH mass (\citealt{Josan09}; \citealt{Carr10}), we only give a qualitative
estimate that the abundance of PBH host UCMHs should be much less
than other UCMHs. For PBH host UMCHs, the X-ray emission from
the accreting baryonic gas flows onto PBHs may totally dominated over dark matter annihilation
within the host UCMH, since the Eddington luminosity is several orders
of magnitude brighter than that of annihilation from the host UCMH, and
the photoionization cross section for hydrogen or helium is much larger
than the Klein-Nishina or pair production cross section for energetic
gamma-rays. It is very difficult for a ``naked'' PBH to reach a sufficient
high accretion rate in the IGM environment,
(\citealt{Barrow79}; \citealt{Carr81}; \citealt{Gnedin95}; \citealt{Miller01};
\citealt{Ricotti07}; \citealt{Ricotti08}; \citealt{Mack08}), but the situation will be quite different if PBHs are
surrounded by UCMHs. The accretion rate and X-ray luminosity of
baryons can change significantly when the
effects of a growth UCMH is involved (\citealt{Ricotti07}; \citealt{Ricotti08}).
However, it is possible that the gas is heated and piles up around the PBH if
the host UCMH is sufficiently massive. Also, the accretion feedback such as
outflows or radiation pressure prevent gas from being totally eaten by
the PBH immediately, if the gas accretion rate significantly exceeds
the Eddington limit. As a consequence, the gas density and temperature within the UCMH
may be significantly higher than the cosmic universal gas
density, and the X-ray emission is totally absorb in the UCMH, but reradiate basically
in the infrared band. In this paper we will give criteria
for X-ray emission escaping from the host UCMH to ionize the IGM.
Also we will compare the importance of X-ray emission from PBH host
UCMHs and dark matter annihilation from total UCMHs in the early
Universe, depending on the abundance of both total UCMHs and PBH
host UCMHs.

Another topic related to the UCMH radiation is that, the formation and evolution history
of the first baryonic structure can be changed by UCMH radiation. Previous
studies showed that the annihilation or decay of the extended distributed
dark matter in the first structure both change the gas temperature and the
chemical properties such as the abundance of molecular
coolants such as H$_{2}$ and HD (\citealt{Biermann06};
\citealt{Stasielak07}; \citealt{RMF07b}). Higher coolants abundance
helps to decrease the gas temperature and favors an early collapse
of the baryon gas inside the halo, but dark matter energy injection
delays this collapsing process. It is still under debate whether
dark matter annihilation or decay inside the dark halo will promote
or suppress the first structure formation. Nevertheless, it is concluded
that either the promotion or suppression effect is quite small for
most dark matter models, as the change of gas temperature in a
virialized halo for various dark matter models is small. If the first large scale dark haloes
contain UCMHs, these UCMHs can inject more annihilation energy into
the halo than the first dark haloes, and potentially play more important role to change
the properties of the first haloes than the extended distributed dark matter in haloes.
Moreover, X-ray emission which comes from PBHs also
suppress the formation of the first baryonic objects. Therefore it is also
worthwhile to study the effects of UCMH radiation on the first structure
formation and evolution.

\begin{table*}
\begin{center}
Notation and definition of some quantities in this paper
\begin{tabular}{|l|l|l|}
\hline
notation & definition & \S/Eq. \\
\hline
$m_{h}(z)$ & UCMH mass at redshift $z$ & \S \ref{secanni}, eq. \ref{halo01}\\
$\rho_{\chi}(r,z)$ & UCMH (density) profile at $z$ & \S \ref{secanni}, eq. \ref{halo02}\\
$R_{h}(z)$ & extent radius of UCMH & \S \ref{secanni}, eq. \ref{halo03}\\
$m_{\chi}$ & DM particle mass & \S \ref{secanni}, eq. \ref{anni04}\\
$\langle\sigma v\rangle$ & DM average anni cross section & \S \ref{secanni}, eq. \ref{anni04}\\
$L_{\rm ann}$ & anni lum of a single UCMH & \S \ref{secanni}, eq. \ref{halo05}\\
$f_{\rm UCMH}$ & $f_{\rm UCMH}(z_{\rm eq})=\Omega_{\rm UCMH}(z_{\rm eq})/\Omega_{\rm DM}$ & \S \ref{secanni}, eq. \ref{ann02}\\
$l_{\rm ann}$ & UCMH anni lum per volume & \S \ref{secanni}, eq. \ref{halo08}\\
$l_{\rm bkgd}$ & homogenous DM anni background  & \S \ref{secanni}, eq. \ref{halo09}\\
$l_{\rm acc}$ & X-ray rad density from PBH host UCMHs & \S \ref{secAccPBH}, eq. \ref{PBH01}\\
$f_{\rm PBH}$ & $\Omega_{\rm PBH}/\Omega_{\rm DM}$ & \S \ref{secAccPBH}, eq. \ref{PBH01} \\
$r_{B}$ & Bondi accretion radius & \S \ref{secAccPBH}, eq. \ref{PBH06}\\
$\mathcal{A}$ & amplification factor of the IGM $T_{\rm m}$ & \S \ref{secAccPBH}, eq. \ref{PBH06}\\
$\dot{m}$ & Dimensional accretion rate & \S \ref{secAccPBH}, eq. \ref{gaccre02}\\
$f_{b}$ & baryonic fraction in a UCMH & \S \ref{secAccPBH}, eq. \ref{gaccre03}\\
$z_{m}$ & charac. redshift for gas accretion & \S \ref{secAccPBH}, eq. \ref{gaccre06}\\
$x_{\rm ion}(z)$ & reionized baryon fraction at $z$& \S \ref{secIonHeat1}, eq. \ref{baseq01}\\
$\epsilon(z)$ & energy deposition rate per volume at $z$ & \S \ref{secIonHeat1}, eq. \ref{bkgd02}\\
$E_{\gamma}$ & photon energy from anni DM & \S \ref{secIonHeat1}, eq. \ref{emission}\\
$E_{X}$ & charac. energy for X-ray emission & \S \ref{secIonHeat1}, ------ \\
$T_{\rm m}$ & IGM gas temperature & \S \ref{secIonHeat1}, eq. \ref{heat01}\\
$f_{\rm H_{2}}$ & molecular hydrogen fraction in IGM gas & \S \ref{secIonHeat1}, eq. \ref{heat04}\\
$L_{\rm halo}$ & total anni lum from a dark halo & \S \ref{secStruc01}, ------  \\
$L_{\rm UCMH}$ & anni lum from UCMHs inside a halo & \S \ref{secStruc01}, ------ \\
$L_{\rm ext}$ & anni lum from extended DM in a halo & \S \ref{secStruc01}, ------ \\
$\epsilon_{\rm loc,iso}$ & energy deposited by isothermal DM halo anni & \S \ref{secStruc01}, eq. \ref{structure01}\\
$\epsilon_{\rm loc,UCMH}$ & energy deposited by UCMHs anni inside a halo & \S \ref{secStruc01}, eq. \ref{deposition2}\\
$\epsilon_{\rm loc,acc}$ & energy deposited by X-rays inside a halo & \S \ref{secStruc02}, eq. \ref{structure02}\\
$z_{m}^{\rm halo}/z_{m}^{\rm bkgd}$ & $z_{m}$ for PBH host UCMHs inside/outside a halo & \S \ref{secStruc02}, ------\\
$\epsilon_{\delta}(z)$ & energy deposition rate density by $\delta$-func SED & \S \ref{SecDis03}, eq. \ref{dis022}\\
factor & ratio of rad of differently distributed UCMHs & \S \ref{SecDis04}, eq. \ref{ratio01}\\
\hline
\end{tabular}
\caption{In this table ``DM'', ``anni'', ``lum'', ``charac.'',
``rad'', ``func'' are short for dark matter, annihilation,
luminosity, characteristic, radiation and function. The subscript
``acc'' is for X-ray emission because X-rays are emitted by gas
accretion onto PBHs.}
\end{center}
\end{table*}\label{notation}

This paper is organized as follows. In Section \ref{secRadfromU}, we calculate the dark
matter annihilation luminosity from UCMHs, and the X-ray emission from
PBH gas accretion. We emphasize on the importance of UCMH annihilation
compared to the homogenous dark matter background annihilation,
and focus on the physical reasons that whether and when the X-ray emission
from PBHs becomes more important than the UCMH annihilation in the early Universe.
In Section \ref{secIonHeat} we discuss the gas heating and
ionization process by the two types of UCMH radiation from
$z\sim1000$ to 10, and investigate the impact of UCMH radiation on the IGM evolution.
Next in Section \ref{secStruc} we show the influences of
UCMH radiation on the first baryonic structure formation and
evolution. The main results of this paper are given in Sections \ref{secIonHeat} and \ref{secStruc}.
In Section \ref{SecDis} we discuss the importance of
UCMHs in the reionization era, the effects of a single massive UCMH in the
first baryonic structure, as well as other secondary effects. Reader could skip
this section and directly go to Section \ref{SecCon}, which presents the
conclusions. In this paper we do not consider dark matter decay, which should have similar
consequences as the annihilation process. Also we fix the
annihilation cross section to be thermal relic $\langle\sigma
v\rangle=3\times 10^{-26}$ cm$^{3}$ s$^{-1}$, although much larger
cross section $\langle\sigma v\rangle=3\times 10^{-24}$ cm$^{3}$
s$^{-1}$ to $10^{-20}$ cm$^{3}$ s$^{-1}$ is proposed in the hope of
explaining the reported Galactic cosmic ray anomalies as the results
of dark matter annihilation (e.g., \citealt{Chang08};
\citealt{Abdo09}; \citealt{Aharonian08}). A larger cross section
with the same dark matter particle mass $m_{\chi}$ can have higher
luminosity and more significant influence on ionizing and heating
the early Universe. Table \ref{notation} gives the notation and
definition of some quantities in this paper.

\section{Radiation from UCMHs}\label{secRadfromU}

In this section we discuss two types of energy emission from UCMHs
in the early Universe: dark matter annihilation, and X-ray emission
from the accreting baryonic gas onto PBHs. As mentioned in Section 1,
the second type of emission is related to a small
fraction of UCMHs, which host PBHs. Generally we still call the second type of energy emission as
UCMH radiation, that is because a PBH is always located in the
center of its host UCMH and belong to a PBH-UCMH system.

\subsection{Dark Matter Annihilation}\label{secanni}

Dark matter annihilation luminosity of nearby UCMHs ($z=0$) is
calculated recently (\citealt{Scott09}; \citealt{Lacki10};
\citealt{Josan10}; \citealt{Saito11}). We assume that UCMHs stop growing
at $z\approx10$ when the structure formation
progressed deeply to prevent dark matter from further accreting. Now we calculate the annihilation
luminosity as the function of redshift in the early Universe before
$z\approx10$, and compare the result to the homogenous
background annihilation. The mass of the UCMHs accreted by dark
matter radial infall is given by (\citealt{Mack07};
\citealt{Ricotti09}; \citealt{Scott09}; \citealt{Josan10})
\begin{equation}
m_{h}(z)=\delta m\left(\frac{1+z_{\rm eq}}{1+z}\right),\label{halo01}
\end{equation}
where $z_{\rm eq}\approx3100$ is the redshift of matter-radiation
equality, and $\delta m$ is the mass of initial dark matter
overdensity. The density profile in an UCMH $\propto r^{-\alpha}$ can be
written as
\begin{equation}
\rho_{\chi}(r,z)=\frac{(3-\alpha)m_{h}(z)}{4\pi R_{h}^{3-\alpha}r^{\alpha}},\label{halo02}
\end{equation}
where the factor $(3-\alpha)/4\pi$ in equation (\ref{halo02}) is obtained by normalizing the
total mass inside the maximum halo extent radius $R_{h}$ as
$\delta m$, and $R_{h}$ is calculated by
\begin{equation}
R_{h}(z)\approx0.019\,{\rm pc}\left(\frac{1000}{1+z}\right)\left(\frac{m_{h}(z)}{M_{\odot}}\right)^{1/3}.\label{halo03}
\end{equation}

\begin{figure}
\centerline{\includegraphics[width=10cm, height=5cm]{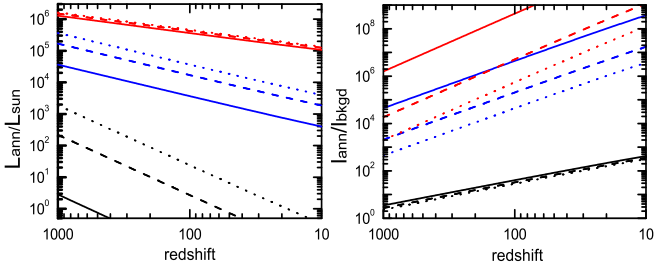}}
\caption{Left: UCMH dark matter annihilation luminosity with the halo profile
$\alpha=1.5$ (black lines), 2.25 (blue lines), 2.9 (red lines), and dark matter particle
mass $m_{\chi}c^{2}$=100 GeV (solid lines), 1 GeV (dashed lines), 100 MeV (dotted lines). We adopt $\delta m=1M_{\odot}$ in this figure.
Right: ratio of UCMH luminosity to homogenous dark matter background annihilation, where
we take the fraction of UCMH in the total dark matter as $f_{\rm UCMH}=10^{-4}$ at $z=z_{\rm eq}$,
and the lines as the same as in left panel.}\label{figratio}
\end{figure}

Dark matter annihilation reduces the density in the inner region of
an UCMH, and makes the density in this region to be flat. Following
\cite{Ullio02} the UCMH power-law density
distribution is truncated at the maximum density
\begin{equation}
\rho(r_{\rm cut})=\rho_{\rm max}=\frac{m_{\chi}}{\langle\sigma v\rangle(t-t_{\rm i})},\label{anni04}
\end{equation}
where $t\approx\frac{2}{3}(1+z)^{-3/2}(\Omega_{\rm
m,0})^{-1/2}H_{0}^{-1}$ is the age of the Universe at a certain
redshift $z$, and $t_{\rm i}\approx$ 77 kyr is the initial age at
$z_{\rm eq}$. Thus the total dark matter annihilation luminosity within the UCMH
can be calculated as
\begin{eqnarray}
L_{\rm ann}&=&\int_{0}^{R_{h}}2\pi r^{2}n_{\chi}^{2}(r)\langle\sigma v\rangle m_{\chi}c^{2}dr\nonumber\\
&=&\frac{2\pi c^{2}}{3}\left(\frac{2\alpha}{2\alpha-3}\right)\textit{K}^{\frac{3}{\alpha}}\langle\sigma v\rangle^{\frac{3-\alpha}{\alpha}}(1+z)^{\frac{9-4\alpha}{\alpha}}\nonumber\\
&&\times\delta m(t-t_{\rm i})^{\frac{3-2\alpha}{\alpha}}m_{\chi}^{\frac{\alpha-3}{\alpha}},
\label{halo05}
\end{eqnarray}
where $\textit{K}=(3-\alpha)(4.66\times10^{8})^{\alpha-3}(1+z_{\rm
eq})^{\alpha/3}(4\pi)^{-1}$ cgs units,
$n_{\chi}(r)=\rho_{\chi}(r)/m_{\chi}$ is the dark matter particle
number density. The UCMH density profile can change from a steep
slope $\alpha=3$ (\citealt{Mack07}) for the
outer part region to $\alpha=1.5$ (\citealt{Bert85}) for the
inner part region, if there is a PBH in the center of the UCMH. In particular,
radial infall onto a central extended overdensity shows a profile
$\rho\propto r^{-9/4}$, which is more widely used as the typical
density profile for most region in UCMHs (\citealt{Ricotti09};
\citealt{Scott09}; \citealt{Josan10}). Taking $\alpha=9/4$, we have
\begin{equation}
L_{\rm ann}=36.3L_{\odot}\langle\sigma v\rangle_{s}^{1/3}m_{\chi,100}^{-1/3}
(1+z)\left(\frac{\delta m}{M_{\odot}}\right),\label{ann01}
\end{equation}
where $\langle\sigma v\rangle_{s}=\langle\sigma
v\rangle/3\times10^{-26}$ cm$^{3}$ s$^{-1}$ and
$m_{\chi,100}=m_{\chi}c^{2}/100$ GeV. According to equation
(\ref{ann01}), the annihilation luminosity decreases with the
evolution of the Universe, basically because the annihilation flats the
inner density profile as showed in equation (\ref{anni04}). The left panel of Fig. \ref{figratio} gives
UCMH annihilation luminosity with different halo profile
$\alpha$ and dark matter particle mass $m_{\chi}$. Note that the
annihilation luminosity can be much brighter for lighter dark matter
particles, and a shallower density profile reduces $L_{\rm ann}$
significantly.\footnote{In Fig. \ref{figratio} the
annihilation luminosities following the density profile equation (\ref{halo02})
are somewhat overestimated for a steep profile $\alpha\sim3$,
because in this case the halo mass within the
truncated radius $r_{\rm cut}$ can no more be neglected. Thus the
normalization factor of the density profile is $\propto[\ln(R_{h}/r_{\rm cut})]^{-1}$,
which is different from the factor $(3-\alpha)/(4\pi)$ in equation (\ref{halo02}).
However, as we show that $L_{\rm ann}$ for a steeper UCMH profile leads to several
orders of magnitude higher $L_{\rm ann}$ than that with
$\alpha=2.25$, the conclusion that a steeper profile gives a
brighter annihilation will not be changed too much even in the limit
case $\alpha=3$. More details of the UCMH profile are discussed in
Section \ref{SecDis02}.} In the limit case
$\alpha\simeq1.5$ or 3, we have $L_{\rm ann}\propto\langle\sigma
v\rangle/m_{\chi}$ or $L_{\rm ann}$ to be independent with
$m_{\chi}$ and $\langle\sigma v\rangle$.
The abundance of UCMH as the function of redshift is still uncertain
today. It can be presented by a parameter
\begin{equation}
f_{\rm UCMH}(z)=\Omega_{\rm UCMH}(z)/\Omega_{\rm DM},\label{ann02}
\end{equation}
where $\Omega_{\rm UCMH}$ and $\Omega_{\rm DM}$ are the comoving abundances of
UCMH and total dark matter with $\Omega_{\rm UCMH}(z)=\Omega_{\rm UCMH}(z_{\rm eq})(1+z_{\rm eq})/(1+z)$
and $\Omega_{\rm DM}(z)=\Omega_{\rm DM}(z=0)$.
We have $f_{\rm UCMH}(z=0)\sim f_{\rm UCMH}(z_{\rm eq})(1+z_{\rm
eq})/(1+10)\sim3\times10^{2}f_{\rm UCMH}(z_{\rm eq})$, which shows that
the UCMH mass grows by up to two order of magnitude from $z_{\rm eq}$ to $z\sim10$. From now on
we take $f_{\rm UCMH}$ as $f_{\rm UCMH}(z_{\rm eq})$ for short to
show the initial abundance of UCMH at matter-radiation equality,
and in current stage $f_{\rm UCMH}$ is taken as a parameter for simplicity.

The mean free path of gamma-ray photons from an UCMH with energy
$E_{\gamma}$ is written as
\begin{equation}
\lambda_{\rm UCMH}=\frac{1}{n_{A}(z)\sigma(E_{\gamma})}\gg\frac{1}{n_{A}(z)\sigma_{T}}\approx3\times10^{3}\;\textrm{pc}\;\left(\frac{1000}{1+z}\right)^{3},
\label{halo06}
\end{equation}
where $n_{A}(z)=n_{A}(1+z)^{3}$ is the atomic number density at redshift $z$.
The average distance of inter-UCMH is estimated as
\begin{eqnarray}
d_{\rm UCMH}&=&\left[\frac{1}{n_{\rm UCMH}(z)}\right]^{1/3}=\left[\frac{\langle\delta m\rangle}{f_{\rm UCMH}\rho_{\rm DM}(z)}\right]^{1/3}\nonumber\\
&\sim&7\;\textrm{pc}\;f_{\rm UCMH,-4}^{-1/3}\left(\frac{\langle\delta m\rangle}{M_{\odot}}\right)^{1/3}\left(\frac{1000}{1+z}\right).
\label{halo07}
\end{eqnarray}
We have the mean free path exceed the inter-UCMH distance
$\lambda_{\rm UCMH}\gg d_{\rm UCMH}$ except for an extremely small
$f_{\rm UCMH}\ll10^{-12}$. Therefore the cosmic UCMH annihilation
also gives an uniform gamma-ray background radiation
field as well as that produced by the homogenous dark matter. UCMH
annihilation luminosity per volume is given by
\begin{eqnarray}
&&l_{\rm ann}=\frac{L_{\rm ann}}{\delta m}f_{\rm UCMH}(z_{\rm eq})\rho_{\rm DM}(z=0)(1+z)^{3}\nonumber\\
&&=1.4\times10^{-28}\,\textrm{erg}\,\textrm{cm}^{-3}\,\textrm{s}^{-1}
\,\langle\sigma v\rangle_{s}^{1/3}m_{\chi,100}^{-1/3}f_{\rm UCMH}(1+z)^{4}.\nonumber\\
\label{halo08}
\end{eqnarray}
On the other hand, the energy injection rate by the
self-annihilation of the homogenous dark matter background per volume is
\begin{eqnarray}
&&l_{\rm bkgd}=\langle\sigma
v\rangle\frac{\rho_{\chi}^{2}c^{2}}{2m_{\chi}}\nonumber\\
&&=3.2\times10^{-43}\,\textrm{erg}\,\textrm{cm}^{-3}\,\textrm{s}^{-1}
\,\langle\sigma v\rangle_{s}m_{\chi,100}^{-1}(1+z)^{6}.
\label{halo09}
\end{eqnarray}
We compare the radiation between UCMH and normal dark matter
annihilation as
\begin{equation}
\frac{l_{\rm ann}}{l_{\rm bkgd}}=4.5\times10^{12}\langle\sigma v\rangle_{s}^{-2/3}
m_{\chi,100}^{2/3}
\left(\frac{10}{1+z}\right)^{2}f_{\rm UCMH}\gg1,
\label{halo10}
\end{equation}
If $f_{\rm UCMH}\geq2.2\times10^{-15}\langle\sigma
v\rangle_{s}^{2/3} m_{\chi,100}^{-2/3}(1+z)^{2}$, the gamma-ray
background due to dark matter annihilation is dominated by
UCMH annihilation. More details depending on the density profile
$\alpha$ and dark matter $m_{\chi}$ can be seen in the right panel
of Fig. \ref{figratio}.

\subsection{Gas Accretion onto PBHs}\label{secAccPBH}

The abundance of PBHs $\Omega_{\rm PBH}$ as a fraction of total dark
matter $\Omega_{\rm DM}$ at $z<z_{\rm eq}$ can be parameterized as
$f_{\rm PBH}=\Omega_{\rm PBH}/\Omega_{\rm DM}$. We ignore
the PBH growth and take $f_{\rm PBH}$ as a constant in the
matter-dominated Universe for two reasons. The first reason is that,
as the accretion processes had been significantly suppressed before
$z\sim10$ due to the relative motion between PBHs and baryon gas,
the PBH growth timescale is $t_{\rm growth}\sim t_{\rm
Salp}\simeq5\times10^8$ yr just reaches or is longer than the universe
age at $t(z\sim10)\sim5\times10^8$ yr. The second reason is that low
mass PBHs has lower accretion rate while high mass PBHs are inclined to
produce outflows, which further increases the accretion timescale
and makes PBH growth to be negligible compared to its host UCMH growth.
A very similar statement to keep a constant $f_{\rm PBH}$ was also proposed in \cite{Ricotti08}.
Keep in mind the PBH abundance $f_{\rm PBH}$ is different from the
UCMH initial abundance $f_{\rm UCMH}$ at $z_{\rm eq}$ as mentioned in
Section \ref{secanni}, because a large amount of UCMH seeds at
$z_{\rm eq}$ should be the initial primordial dark matter overdensity
but not PBHs. According to the density primordial perturbation
theory, generally we have the relation $f_{\rm PBH}\ll f_{\rm
UCMH}$, which will be discussed in details in Section
\ref{secAccPBH1}.

Much work has been done to show the effects of radiation from PBH
or early black hole accretion on the early Universe thermal and ionization
history (e.g., \citealt{Barrow79}; \citealt{Gnedin95};
\citealt{Miller01}; \citealt{Ricotti07}; \citealt{Ricotti08};
\citealt{RMF08}). Our goal in this section is to focus on the
importance of PBH gas accretion radiation compared to the overall
UCMH dark matter annihilation. The X-ray emission form accreting
PBHs may lead to a very different heating and ionization history
of the early Universe compared to the dark matter annihilation. The
X-ray luminosity from an individual PBH with mass $M_{\rm PBH}$ can
be written as $\eta L_{\rm Edd}=4\pi \eta Gm_{p}M_{\rm
PBH}/\sigma_{T}c\simeq3.3\times10^{3}\eta_{-1} L_{\odot}(M_{\rm
PBH}/M_{\odot})$ with $L_{\rm Edd}$ and $\eta_{-1}=\eta/0.1$ being
the Eddington luminosity and average radiation efficient of all PBHs
respectively. This X-ray luminosity is much higher than the dark
matter annihilation luminosity equation (\ref{ann01}). Thus the
X-ray radiation density in the early Universe $z>10$ can be written
as
\begin{eqnarray}
l_{\rm acc}&=&\left(\frac{L_{\rm Edd}}{M_{\rm PBH}}\right)\eta f_{\rm PBH}\rho_{\rm DM}(z=0)(1+z)^{3}\nonumber\\
&\simeq&1.3\times10^{-26}\,\textrm{erg}\,\textrm{cm}^{-3}\,\textrm{s}^{-1}\,\eta_{-1} f_{\rm PBH}(1+z)^{3}.\label{PBH01}
\end{eqnarray}
Combing equations (\ref{halo08}) and (\ref{PBH01}), the ratio
between PBH accretion luminosity and UCMH dark matter annihilation
luminosity is
\begin{equation}
\frac{l_{\rm acc}}{l_{\rm ann}}\simeq9\langle\sigma v\rangle_{s}^{-1/3}m_{\chi,100}^{1/3}
\left(\frac{\eta_{-1}f_{\rm PBH}}{f_{\rm UCMH}}\right)\left(\frac{10}{1+z}\right).\label{PBH02}
\end{equation}
Since the IGM heating rate due to the energy injection by PBH X-ray
emission or UCMH dark matter annihilation is proportional to both
the energy injection rate, and the IGM cross section for all the
interactions suffered by the UCMH emitted photons in X-ray band
$E_{X}$ or annihilation emitted gamma-ray photons $E_{\gamma}$, the
importance of the IGM gas heating by X-ray emission and UCMH dark
matter annihilation can be estimated by the ratio $l_{\rm
acc}\sigma_{\rm tot}(E_{X})/l_{\rm ann}\sigma_{\rm tot}(E_{\gamma})$
with $\sigma_{\rm tot}$ labeling the total cross sections in
different photon energy range. If we roughly take the X-ray and IGM
interaction cross section $\sigma_{\rm tot}(E_{X})$ the Thomson
cross section, and the high energy photon interaction $\sigma_{\rm
tot}(E_{\gamma})$ the Klein-Nishina cross section (see Section
\ref{secIonHeat1} for more accurate calculations), the energy
deposition in the IGM due to gas accretion $\epsilon_{\rm acc}$ and
annihilation $\epsilon_{\rm ann}$ radiation is estimated as
\begin{equation}
\frac{\epsilon_{\rm acc}}{\epsilon_{\rm ann}}\simeq3.2\times10^{5}
\langle\sigma v\rangle_{s}^{-1/3}m_{\chi,100}^{4/3}
\left(\frac{\eta_{-1}f_{\rm PBH}}{f_{\rm UCMH}}\right)\left(\frac{10}{1+z}\right),\label{PBH03}
\end{equation}
which gives the first conclusion that the X-ray heating may become
totally dominated over dark matter annihilation in the early
Universe if the PBH abundance exceeds a critical value as
\begin{equation}
\frac{\eta f_{\rm PBH}}{f_{\rm UCMH}}\geq3.1\times10^{-7}\langle\sigma
v\rangle_{s}^{1/3}m_{\chi,100}^{-4/3}\left(\frac{1+z}{10}\right).\label{PBH04}
\end{equation}

Theoretically the value of $\eta f_{\rm PBH}/f_{\rm UCMH}$ includes
many uncertainties. In general, there are at least three reasons to have
a low value $\eta f_{\rm PBH}/f_{\rm UCMH}\ll1$: density perturbation scenarios prefer low
initially value of $f_{\rm PBH}/f_{\rm UCMH}$; inefficient radiation
$\eta\ll1$ is favored by low mass PBHs while accretion feedback
decreases $\eta$ for high mass PBHs or PBHs with high mass UCMHs; and X-ray emission from PBHs
can be trapped inside the surrounding host UCMHs which
accumulate baryons.

\subsubsection{PBH Abundance}\label{secAccPBH1}

Either PBHs or UCMH overdensity seeds are produced basically by the
density perturbation in the very early Universe during some special
epoches such as inflation or phase transitions. The cosmological
abundance of UCMHs can be estimated by integrating from the
overdensity seed threshold $\sim10^{-3}$, to the PBH formation threshold
$\delta_{c}\sim1/3$ (\citealt{Ricotti09}; \citealt{Scott09}). Similarly, the PBH abundance is estimated by
integrating the perturbation above $\delta_{c}\sim1/3$
(\citealt{Green97a, Green97b}). Assuming a Gaussian perturbation at
a formation redshift $z_{f}\gg z_{\rm eq}$ to produce both PBHs and
UCMH seeds, the ratio $f_{\rm PBH}/f_{\rm UCMH}$ at matter-radiation equality can be directly
traced back to formation time $z_{f}$ (\citealt{Carr10};
\citealt{Khlopov10}). As a result, the relative abundance of PBHs to
UCMH overdensity seeds formed at redshift $z_{f}$ is written
as
\begin{eqnarray}
\frac{f_{\rm PBH}}{f_{\rm UCMH}}&=&\frac{\int_{\delta_{c}}^{1}\exp\left[-\frac{\delta^{2}}{2\sigma(z_{f})^{2}}\right]d\delta}
{\int_{10^{-3}}^{\delta_{c}}\exp\left[-\frac{\delta^{2}}{2\sigma(z_{f})^{2}}\right]d\delta}\nonumber\\
&\simeq&\exp\left[\frac{10^{-6}-\delta_{c}^{2}}{2\sigma^{2}}\right]\simeq\exp\left(-\frac{1}{18\sigma^{2}}\right).\label{PBH05a}
\end{eqnarray}
The perturbation variance at $z_{f}$ is roughly given by
$\sigma(z_{f})\simeq9.5\times10^{-5}[M_{\rm
hor}(z_{f})/10^{56}\,\textrm{g}]^{(1-n)/4}$ (\citealt{Green97a}),
with $M_{\rm hor}(z_{f})$ and $n$ being the horizon mass and mass
spectrum index at $z_{f}$. Taking $n\leq1.3$ (\citealt{Lidsey95}),
the ratio $f_{\rm PBH}/f_{\rm UCMH}$ from a Gaussian perturbation is
the function of horizon mass as
\begin{equation}
\frac{f_{\rm PBH}}{f_{\rm UCMH}}\leq\exp\left[-\left(\frac{M_{\rm hor}(z_{f})}{5.5\times10^{10}
\,\textrm{g}}\right)^{(n-1)/2}\right],\label{PBH05}
\end{equation}
which means the value of $f_{\rm PBH}/f_{\rm UCMH}$ becomes
$\ll1$ for $M_{\rm hor}(z_{f})\gg5.5\times10^{10}$ g, not
to mention the fact that the masses of dark matter overdensity seeds
or PBHs are even lower than the horizon mass $\delta m\ll M_{\rm
hor}(z_{f})$ and $M_{\rm PBH}\ll M_{\rm hor}(z_{f})$. Combing
equations (\ref{PBH03}) and (\ref{PBH05}), X-ray emission from gas
accretion hardly becomes the dominated heating source in the early
Universe, except for low mass PBHs  $M_{\rm PBH}\ll M_{\rm
hor}(z_{f})<3.7\times10^{18}$ g in the Gaussian perturbation scenario.
However, PBHs in this mass range should either have disappeared within a Hubble time
due to the Hawking evaporation, or too small to accrete the IGM gas.

As a result, the initially Gaussian density perturbation at
a certain epoch is not able to generate sufficient abundant PBH to
dominated over the total UCMH dark matter annihilation emission,
basically because the large amplitude part of a Gaussian
distribution is highly suppressed. On the other hand, non-Gaussian
perturbation may give an even lower probability of PBH formation, as
the large fluctuation can be suppressed in the non-Gaussian
distribution and further decrease the ratio of $f_{\rm PBH}/f_{\rm
UCMH}$ (\citealt{Bullock97}).

However, other mechanisms such as different formation epoches for UCMHs and PBHs,
different early inflationary potential, double inflation models, various phase transitions, and
cosmic string collapse may enhance the high amplitude perturbation
and increase the PBH abundance (see \citealt{Khlopov10} and
references therein). Also, it is still arguable if all the
$\delta>10^{-3}$ perturbation could produce dark matter overdensity
in the radiation dominant era. For example, \cite{Ricotti09}
requires the host UCMHs around PBHs to have similar initial
perturbation amplitude as PBHs, while \cite{Scott09} has less strict
requirement as $\delta>10^{-3}$ to form the initial dark matter
overdensity. There are more physical uncertainties to estimate the
abundance of PBHs and UCMHs produced by other mechanisms than a
simple Gaussian distribution assumption. Therefore we still take
$f_{\rm PBH}$ as a free parameter which satisfies $f_{\rm
PBH}\ll f_{\rm UCMH}$ to describe the relative abundances between
PBHs and UCMHs.

\subsubsection{Inefficient Radiation}\label{secAccPBH2}

Another effect to constrain the X-ray luminosity density in the early
Universe by the gas accretion onto PBHs is the low radiation efficiency due
to low accretion rate onto low mass PBHs, or the significant
radiative feedback, thermal outflow and suppressed accretion rate
due to accretion onto high mass PBHs or PBHs with high mass host UCMHs.

In principle the mass distribution of PBHs is broad enough to cover
the range from the Planck mass $\sim10^{-5}$ g to thousands of solar
mass $10^{5} M_{\odot}$ (e.g., \citealt{Carr10}). As mentioned in the above Section
\ref{secAccPBH1}, if PBHs are formed from the Gaussian perturbation
with the variation $\sigma\propto M^{-(n-1)/4}$ and the index $n>1$, low mass
PBHs should be more abundant because of the higher density
perturbation variance $\sigma$ for lower mass $M$. Also, phase
transition models give PBH mass or UCMH seed less than 1 $M_{\odot}$
(\citealt{Scott09}). On the other hand, keep in mind in the IGM
environment a ``naked'' PBH without a host UCMH can never reach the
Eddington accretion rate $\dot{M}_{\rm Edd}^{\rm PBH}=L_{\rm
Edd}/c^{2}\simeq1.4\times10^{17}(M_{\rm PBH}/M_{\odot})$ g s$^{-1}$,
unless its mass is $M_{\rm
PBH}\geq360\,M_{\odot}[1000/(1+z)]^{3/2}$. The surrounding host UCMH
increases the accretion rate if the PBH mass is $M_{\rm
PBH}>100M_{\odot}$ (\citealt{Ricotti08}, their Fig. 4). Note that an ideal case is
$\eta\simeq$min$\{0.1\dot{m},1\}$ after the accretion become
super-Eddington, while the typical accretion efficiency for quasars
or microquasars disk is $\eta\sim0.15$. For low mass PBHs with
$\dot{m}\ll1$ the radiation efficiency is estimated as
$\eta\simeq0.01\dot{m}^{2}$ for the spherical case
(\citealt{Shaprio73a,Shaprio73b}), which gives much lower efficiency
than the high accretion rate that $\eta\ll0.1$.

If high mass PBHs ($100M_{\odot}<M_{\rm PBH}<10^{5}M_{\odot}$)
successfully form with an appreciable abundance compared to the
low mass PBHs, as discussed by some previous authors
(\citealt{Mack07}; \citealt{Saito08}; \citealt{Frampton10}), or the host UCMH seeds are
more massive than the PBHs $\delta m\gg M_{\rm PBH}$ (\citealt{Ricotti09}, more details see
Section \ref{secAccPBH3}), the Bondi accretion rates onto these PBHs with their host UCHMs can
significant exceed the Eddington limit after some critic redshifts
(\citealt{Ricotti08}). However, a spherical super-Eddington
accretion is generally unstable and inclined to drive high mass loss
rate with thermal outflows (e.g., \citealt{Smith06}). Recent
simulations show that radiative feedback may become important to
reduce or even quench the accretion process periodically(\citealt{Milo09a,
Milo09b}; \citealt{Park11}). Also, the thermal heating by the outflow
energy or radiative feedback will increase the temperature of the gas
around PBHs and decrease the Bondi radius and accretion rate onto PBHs.
Besides the spherical accretion case, the falling gas angular
momentum will become important for $\dot{m}\gg1$, and form an
accretion disk around PBHs. However, the physics of the
super-Eddington accretion disks is still not clearly known. Various
types of super-Eddington accretion disk models have been proposed,
such as the optically-thick advection dominated accretion flow
(ADAF, \citealt{Narayan94}, \citealt{Narayan98}), the adiabatic
inflow-outflow (ADIO, \citealt{Blandford99}), the
convection-dominated accretion flow (CDAF, \citealt{Narayan00}), the
``polish doughnuts'' torus (\citealt{Abramo78}) and the thick slim
disk (\citealt{Abramo88}). In most cases the super-Eddington
accretion disk advects most of its heating energy inward into the
black hole without emission, and has a low radiation efficiency $\eta$ for high accretion rate
$\eta<1$ (\citealt{Abramo88}; \citealt{Narayan98} or \citealt{Abramowicz11} for a review).

In a brief summary, low average radiation efficiency $\eta$ in
equations (\ref{PBH01}) to (\ref{PBH03}) are favored because of the
low accretion rate onto the low mass PBHs, and
radiative or viscous feedback and outflows of accretion onto high
mass PBHs or PBHs with high mass host UCMHs, which also leads to a low value of
$\eta f_{\rm PBH}/f_{\rm UCMH}$ and suppress the importance of PBH
X-ray radiation from PBHs compared to the overall UCMH dark matter
annihilation. From now on we consider the X-ray emission is mainly contributed
by the accreting PBHs with $\dot{m}\gg1$.

\subsubsection{Radiation Trapping in Host UCMHs}\label{secAccPBH3}

Some previous works discussed that the accretion flow around PBHs is
Compton thin in most cases, since in the sub-Eddington accretion
case the spherical flow is transparent near the PBH, while in the
super-Eddington accretion case the accretion flow are inclined to form an
accretion disk (e.g., \citealt{Ricotti08}). However, sufficient high
mass UCMHs can accrete and thermalize baryons from the ambient IGM,
\textit{even there are no PBHs in the center of these UCMHs}. As the gravity
potential at the outer edge of the host UCMH is
mainly contributed by the UCMH mass, but the accretion onto the
center PBHs is according to the PBH mass, the accretion rate into
the host UCMHs is not necessarily equal to the accretion
rate onto the center PBHs. In other words, baryons can be
firstly accumulated and virialized inside the host UCMH during
the accretion from the IGM to the inner UCMH region, followed by a secondary
accretion onto the center PBH and feedback (outflow) from the accreting
PBHs. Based on this consideration, the baryons inside the UMCH can
be divided into two components: the piled up baryons inside the
UCMH, and the accretion spherical flow or disk around the center
PBH. Although the optical depth of the accretion gas or disk, which
is mainly contributed by the depth around the inner horizon region
$r_{\rm}\sim R_{\rm Sch}$ is transparent to X-ray photons, the X-ray
emission can still be trapped and absorbed by the piled up baryons
inside the host UCMH, and reradiate photons with much longer wavelength
into the outer IGM environment. Quantitative analysis is given as
follows. Part of the treatment is similar to an analogy discussion on the dark
matter structure formation and baryons filling process (\citealt{Hoeft06}; \citealt{Okamoto08}).

If UCMH dark matter annihilation does not change the IGM
temperature evolution, the IGM temperature is approximately coupled with the cosmic
microwave background (CMB) temperature before the decoupling time $z_{\rm dec}\sim100$,
and the IGM sound speed before $z_{\rm dec}$ is $c_{s}\simeq5.7$ km
s$^{-1}\left(\frac{1+z}{1000}\right)^{1/2}$. In general, the UCMH annihilation heating and PBH emission without trapping
increases the IGM temperature. We introduce an amplification factor $\mathcal{A}$ that $T_{\rm m}=\mathcal{A}T_{\rm CMB}$
at $z>z_{\rm dec}$, where $T_{\rm m}$ and $T_{\rm CMB}$ are the temperature of the IGM and CMB respectively, and
$\mathcal{A}$ depends on the UMCH profile and annihilation properties, as we will calculate in Section \ref{secIonHeat}.
The sound speed $c_{s}\propto T^{1/2}$ becomes $c_{s}\simeq5.7$ km
s$^{-1}\mathcal{A}^{1/2}\left(\frac{1+z}{1000}\right)^{1/2}$, and the Bondi accretion
radius (i.e., the accretion sonic sphere) of a PBH-UCMH system at $z>z_{\rm dec}$ is
\begin{equation}
r_{B}\approx\frac{Gm_{h}}{c_{s}^{2}}\approx\frac{400\,\textrm{pc}}{(1+z)^{2}}\mathcal{A}^{-1}\left(\frac{\delta m}{M_{\odot}}\right).\label{PBH06}
\end{equation}
Equation (\ref{PBH06}) is derived under the assumption that the Bondi
radius is larger than the UCMH size $r_{B}>R_{h}$. Furthermore, if
$r_{B}>2R_{h}$, the virial temperature of the host UCMH $T_{\rm
vir}\simeq \left(\frac{\mu m_{p}}{2k_{B}}\right)\left[\frac{G m_{h}(z)}{R_{h}}\right]$ is greater than the temperature
of the ambient IGM gas $T_{\rm vir}>T_{\rm m}$. According to the general virial theorem,
the thermal pressure of the gas due to virialized heating is weak
compared to gravity of the UCMH. In this case we consider the IGM
baryons should fall into the UCMH unimpeded, regardless of the
center PBH mass (\citealt{Hoeft06}; \citealt{Okamoto08}).

The criterion $r_{B}>2R_{h}$ at $z>z_{\rm dec}$ gives
\begin{equation}
\left(\frac{\delta m}{M_{\odot}}\right)>1600\mathcal{A}^{3/2}\left(\frac{1+z}{1000}\right).\label{PBH07}
\end{equation}
Note that higher IGM temperature around UCMH, i.e., higher $\mathcal{A}$ gives a higher minimum UCMH mass
to attract baryons. Similar result can be derived for the case after decoupling
$z<z_{\rm dec}$, where the IGM gas temperature decoupled with the CMB
temperature and dropped adiabatically as $T_{\rm ad}\propto(1+z)^{2}$ without any heating sources.
We still take the factor $\mathcal{A}\geq1$ to measure the IGM temperature increase due to annihilation
$T_{\rm m}=\mathcal{A}T_{\rm ad}$. Then using the criterion $r_{B}>2R_{h}$,
we find that baryons fall into UCMHs unimpeded at $z<z_{\rm dec}$ if
\begin{equation}
\left(\frac{\delta m}{M_{\odot}}\right)>160\mathcal{A}^{3/2}\left(\frac{1+z}{100}\right)^{5/2}.\label{PBH08}
\end{equation}
As a result, if the UCMH initial overdensity seed is $\delta
m>1600\mathcal{A}^{3/2}M_{\odot}$ for $z_{\rm dec}<z<1000$, or $\delta
m>160\mathcal{A}^{3/2}M_{\odot}$ for $z<z_{\rm dec}$, the IGM gas can always fill into the
UCMH \textit{no matter it includes a PBH or not}. Otherwise for a lower $\delta m$, the
critical redshift $z_{c}$ below which the UCMH accrete is
$(1+z_{c})<0.63\mathcal{A}^{-3/2}(\delta m/M_{\odot})$ for $z>z_{\rm dec}$ and
$(1+z_{c})<13\mathcal{A}^{-3/5}(\delta m/M_{\odot})^{2/5}$ for $z<z_{\rm dec}$. If the
UCMH hosts a PBH in the center, baryons are still able to piles
up and thermalized in the host UCMH due to the gas virialization.

The lower bound of gas accretion rate into the UCMH can be estimated
as
\begin{eqnarray}
\dot{M}_{\rm UCMH}&=&4\pi r_{B}^{2}c_{s}\rho_{\rm gas}(z)
>4\pi R_{h}^{2}v_{ff}(R_{h})m_{b}n_{b}(z)\nonumber\\
&\sim&8.2\times10^{16}\,\textrm{g}\,\textrm{cm}^{-3}(1+z)^{1/2}\left(\frac{\delta m}{M_{\odot}}\right),\label{gaccre01}
\end{eqnarray}
where $v_{ff}(R_{h})$ is the free fall velocity at $R_{h}$. If all
the gas into the UCMH is totally accreted onto the center PBH, the
dimensionless accretion rate of the PBH is
$\dot{m}=\dot{M}_{g}/\dot{M}_{\rm Edd}^{\rm PBH}$ is
\begin{equation}
\dot{m}>18.0\left(\frac{1+z}{1000}\right)^{1/2}\left(\frac{\delta m}{M_{\rm PBH}}\right)>1,\label{gaccre02}
\end{equation}
with $\dot{M}_{\rm Edd}^{\rm PBH}$ being the Eddington limit accretion rate
onto the central PBH. Note that the ideal $\dot{m}$ can be even higher if the initial host UCMH
is more massive than the center PBH $\delta m\gg M_{\rm PBH}$
as discussed in \cite{Ricotti09}. However, the real accretion rate should be
less than the value in equation (\ref{gaccre02}) for two reasons. First, the baryons can be heated and virialized
during the accretion process in the UCMH, and has a temperature $\sim T_{\rm vir}$
warmer than the IGM $T_{\rm m}$ to increase the gas pressure and
decrease the accretion rate onto the PBH. And the gas temperature is further increased $\gg T_{\rm vir}$ near
the PBH due to PBH emission and ionization. Also,
super-Eddington accretion disks are also inclined to drive outflows.
The positive Bernoulli parameter over most of the ADAFs due to the
small radiation loss may trigger strong outflows or jets
(\citealt{Narayan94}), and produce an ADIOs in which outflow carries
away most of flow mass and energy (\citealt{Blandford99}). Also, CDAFs may
produce a ``convective envelope'' with no accretion onto the black hole
(\citealt{Narayan00}). In general accretion disks with
super-Eddington accretion rate are inevitably accompanied by
outflows and winds, which significantly decrease the final accretion
rate onto the black hole. In the PBH case, these outflows should be injected
back to the host UCMH environment.

The upper bound of the baryonic fraction in the UCMH is the
universal fraction $\Omega_{\rm b}/\Omega_{\rm m}$. However, as the
UCMH grows following equation (\ref{halo01}), we adopt a more
conservative method to estimate the lower bound of baryon fraction
$f_{b}$ inside the UCMH. We estimate the baryonic fraction in the
UCMH $f_{b}$ (the mass ratio between gas and dark matter) as
\begin{equation}
(\dot{M}_{\rm UCMH}-\dot{M}_{\rm PBH})(t-t_{\rm i})\sim m_{h}(z)f_{b}\label{gaccre03}
\end{equation}
where we take $\dot{M}_{\rm UCMH}\gg\dot{M}_{\rm PBH}$, i.e., most of the accreted gas into the UCMH is piled up
without being immediately eaten by the PBH. Combining equations
(\ref{gaccre01}) and (\ref{gaccre03}), we have the lower bound of baryon fraction
to be $f_{b}\geq7.6\times10^{-3}$, which is a constant independent of the
redshift.

The optical depth of the piled up gas in the UCMH due to Compton
scattering is
\begin{eqnarray}
\tau&\sim&x_{e}\sigma_{T}\int\frac{\rho_{\chi}(r)f_{b}}{\mu m_{p}}dr.\label{gaccre041}
\end{eqnarray}
Since $f_{b}$ from equation (\ref{gaccre03}) is a constant, and UCMH growth does not
change the steep region $\rho_{\rm DM}\propto r^{-\alpha}$ with $\alpha=9/4$ but only increase $R_{h}$
and flats the region $r<r_{\rm cut}$ (see equation [\ref{anni04}]), we take the baryon fraction to be uniformly distributed
in the UCMH, both in the steep and flat region. Therefore the column density of the baryon gas inside the UCMH depends on
the UCMH profile, which depends on the dark matter properties ($\langle\sigma v\rangle, m_{\chi}$)
given by equation (\ref{anni04}). Actually the baryon profile can be steeper in the flat region
of the halo $r<r_{\rm cut}$ since the dark matter annihilation flats the
inner halo profile, thus gives an even larger optical depth. Furthermore, we consider the baryon gas is
ionized $x_{e}\sim1$ inside the UCMH, at least in the flat region $r<r_{\rm cut}$.
We check that the Str\"{o}mgren radius of the PBH emission $r_{S}$ satisfies $r_{\rm cut}<r_{S}< R_{h}$,
as the heated gas near the PBH can reach a temperature as high as the Compton temperature $\sim10$ keV in the ionized region.
The hot ionized gas around the PBH produces a small sonic sphere in the dense baryon region near the PBH,
decreases the accretion rate onto the PBH, giving $\dot{M}_{\rm UCMH}\gg\dot{M}_{\rm PBH}$
as mentioned in equation (\ref{gaccre03}). Note that there should be two distinct sonic spheres, the sphere for the host UCMH outside $R_{h}$,
and that for the center PBH inside the UCMH. This scenario is similar with \cite{Wang06} that
an accreting BH has two Bondi spheres, a smaller inner sphere in the hot gas region and a larger one in the outer cooler region.
Therefore we take $x_{e}\sim1$ in equation (\ref{gaccre041}). The optical
depth is written as
\begin{eqnarray}
\tau&\sim&\left(\frac{\sigma_{T}f_{b}}{\mu m_{p}}\right)\int\rho_{\chi}(r)dr\nonumber\\
&\geq&10\left(\frac{\delta m}{M_{\odot}}\right)^{1/3}\left(\frac{1+z}{1000}\right)^{5/6}m_{\chi,100}^{5/9}
\langle\sigma v\rangle_{s}^{-13/9},\label{gaccre04}
\end{eqnarray}
Hereafter we take $\langle\sigma v\rangle_{s}=1$.
Combining equations (\ref{PBH07}) and (\ref{gaccre04}), we conclude
that before the decoupling $z>z_{\rm dec}$ the gas is always Compton
thick to the X-ray emission from the center PBH accretion when the host
UCMH itself accretes baryons. After the decoupling $z<z_{\rm dec}$
the redshift range that X-ray emission escapes is
\begin{equation}
\left\{
\begin{array}{l}
(1+z)<13\left(\frac{\delta m}{M_{\odot}}\right)^{2/5}\mathcal{A}^{-3/5} \quad\quad\quad \left(\frac{\delta m}{M_{\odot}}\right)<7m_{\chi,100}^{-5/6}\mathcal{A}^{3/4}\\
(1+z)<62\left(\frac{\delta m}{M_{\odot}}\right)^{-2/5}m_{\chi,100}^{-2/3} \quad\quad\; \left(\frac{\delta m}{M_{\odot}}\right)>7m_{\chi,100}^{-5/6}\mathcal{A}^{3/4}
\end{array}
\right.\label{gaccre05}
\end{equation}
From equation (\ref{gaccre05}) there is a maximum $z_{m}$ in the
case of $r_{B}>2R_{h}$ that
\begin{equation}
z_{m}^{(r_{B}>2R_{h})}=29m_{\chi,100}^{-1/3}\mathcal{A}^{-3/10}-1,\label{gaccre06}
\end{equation}
if $z>z_{m}$, X-ray photons will be totally trapped. Note that $z_{m}$
insensitively decreases with the increasing of the IGM temperature factor $\mathcal{A}$.
A larger optical depth due to a deeper baryon profile at $r<r_{\rm cut}$ gives an even lower $z_{m}$.
Also, the range of $\delta m$ applied in equation (\ref{gaccre05}) is
\begin{equation}
0.64M_{\odot}\mathcal{A}^{3/2}<\delta m<80M_{\odot}m_{\chi,100}^{-5/3}.\label{gaccre05a}
\end{equation}
In other words, in the $r_{B}>2R_{h}$ case,
no X-rays can escape the host UCMH if $\delta m\geq80M_{\odot}m_{\chi,100}^{-5/3}$.
More massive PBHs are easier to reach Eddington accretion rate, but more difficult to
produce a transparent baryon environment in the host UCMHs.

On the other hand, if $r_{B}<2R_{h}$ (i.e., $T_{\rm vir}(R_{h})<T_{\rm m}$), most part of the UCMH gravity
potential well is not deep enough to compress the gas and overcome
the pressure barrier of the gas virialization heating. In this case
the UCMH itself cannot accrete and thermalize baryons except for the region in
the radius $r_{B}'$ where satisfies $[Gm_{h}(r\leq
r_{B}')/(2r_{B}')]=c_{s}^{2}$. At $z>z_{\rm dec}$, the critical
radius $r_{B}'$ for a pure UCMH profile $\rho_{\chi}\propto
r^{-\alpha}$ is
\begin{equation}
r_{B}'=R_{h}\left(\frac{Gm_{h}}{2c_{s}^{2}R_{h}}\right)^{1/(\alpha-2)},\label{gaccre06a}
\end{equation}
where we apply $\alpha=9/4$ from equation (\ref{halo05}). The part
of UCMH inside $r_{B}'$ can accrete and heat baryons. Similar to
equations (\ref{gaccre01}) to (\ref{gaccre04}), the accretion rate
into the region $r\leq r_{B}'$ in the unit of Eddington accretion
rate of the center PBH at $z>z_{\rm dec}$ is
\begin{eqnarray}
\dot{m}&=&\frac{2\pi r_{B}'^{2}c_{s}m_{b}n_{b}(z)}{\dot{M}_{\rm Edd}^{\rm PBH}}\nonumber\\
&\approx&3.9\times10^{-11}\mathcal{A}^{-15/2}\left(
\frac{100}{1+z}\right)^{9/2}\left(\frac{\delta m}{M_{\odot}}\right)^{5}\left(\frac{\delta m}{M_{\rm PBH}}\right),\label{gaccre07}
\end{eqnarray}
where the factor $2 \pi$ is due to the suppressed accretion at $r>r_{B}'$ in the UCMH,
thus the baryon density is half of the ambient gas density. Assuming $\delta m=M_{\rm PBH}$,
equation (\ref{gaccre07}) shows that only high mass PBHs ($M_{\rm PBH}>100M_{\odot}$) are able
to produce super-Eddington accretion if there is no accretion feedback. This is
basically consistent with the results in \cite{Ricotti08}. However,
we should mention two things. The first thing is that, if $\delta m>M_{\rm PBH}$, the ideal
accretion rate in equation (\ref{gaccre07}) also increases. The second thing, which
is similar to the analysis below equation (\ref{gaccre02}) is that, the real accretion rate onto the PBH is lower
than the ideal $\dot{m}$ due to higher gas temperature and accretion feedback.
As a result, we find that baryons can be accumulated and virialized inside the UCMH
region $r\leq r_{B}'$ with a baryon fraction approximately as
$f_{b}\sim10^{-3}\mathcal{A}^{-9/2}(1+z)^{-3}(\delta m/M_{\odot})^{3}$. Using the
condition $\dot{m}>1$ in equation (\ref{gaccre07}) at $z>z_{\rm dec}$ for a sufficient radiation efficiency,
the baryon optical depth inside radius $r_{B}'$ is
\begin{eqnarray}
\tau&\sim&1.1\times10^{-2}\mathcal{A}^{-9/2}(1+z)^{-13/6}m_{\chi,100}^{5/9}\left(\frac{\delta m}{M_{\odot}}\right)^{10/3}\nonumber\\
&>&5.0m_{\chi,100}^{5/9}\mathcal{A}^{1/2}\left(\frac{M_{\rm PBH}}{\delta m}\right)^{2/3}\left(\frac{1+z}{100}\right)^{5/6}>1,\label{gaccre08}
\end{eqnarray}
where we also take the baryon profile
inside the UCMH is proportional to the dark matter profile for simplicity,
and $x_{e}\sim1$.
According to equation (\ref{gaccre08}), we consider the UCMH is
optically thick to the X-ray emission in the case of $r_{B}<2R_{h}$
and $z>z_{\rm dec}$, unless $\delta m\gg M_{\rm PBH}$ or dark matter
particle mass $m_{\chi,100}\ll1$.

After decoupling $z<z_{\rm dec}$, we find that baryons can be
accumulated and virialized inside a radius
\begin{equation}
r_{B}'\approx0.012\,\textrm{pc}\,\mathcal{A}^{-4}\left(\frac{30}{1+z}\right)^{8}
\left(\frac{\delta m}{M_{\odot}}\right)^{3}\label{gaccre09}
\end{equation}
with a upper bound $z_{m}$ that
\begin{equation}
z_{m}^{(r_{B}'<2R_{h})}\simeq32\mathcal{A}^{-3/8}\left(\frac{\delta m}{M_{\rm PBH}}\right)^{1/2}m_{\chi,100}^{-5/12}-1,\label{gaccre10}
\end{equation}
below which ($z<z_{m}$) the gas is Compton thin
inside radius $r_{B}'$. Higher ratio $\delta m/M_{\rm PBH}\gg 1$ or lighter dark matter
$m_{\chi,100}\ll1$ increases $z_{m}$, which also decreases slightly if annihilation
effect is included to heat the IGM gas ($\mathcal{A}>1$).

Not only the baryons inside UCMHs can trap X-ray photons, but also
the outflows driven by PBH accretion feedback also absorb the X-ray
emission as well. As mentioned before, the super-Eddington accretion
rate onto the PBH is unstable to trigger strong outflows in
both spherical and disk cases. An optically thick ``outflow
envelope'', both in the polar and equatorial region around the PBHs,
forms to cover the PBH and totally or mostly absorb X-ray emission
form the inner accretion flows (\citealt{Igumenshchev03};
\citealt{Kohri05}; \citealt{Poutanen07}; \citealt{Abolmasov09}). In
this case, the Compton heating in the outflow region should be
important to increase the gas pressure and temperature, balance the
gravity well, reemit thermalized photons from outflows, and regular
and the accretion rate onto the PBH (\citealt{Wang06}). It is likely
to have a steady state or periodically changing outflow envelope
covering the whole PBH, but the details are still an open
question which is beyond the purpose of this paper. What we want to show is
that, even though the accretion disk itself around PBH is optical
thin to X-ray radiation, the X-ray emission can be still absorbed in the
outflow envelope due to the accretion feedback and disk instability
in the super-Eddington case, not to mention the optical-thick or
geometry thick disks which absorb X-ray emission by themselves.

We give a summary of Section \ref{secAccPBH}. Equation
(\ref{PBH01}) computes the X-ray radiation density due to baryon
gas accretion onto PBHs. The ratio $\eta f_{\rm PBH}/f_{\rm UCMH}$
is parameterized in this paper. The much lower probability of PBH
formation compared to the UCMH formation, and inefficient radiation
due to low mass PBHs ($\delta m\ll 100M_{\odot}$) or accretion
feedback from high mass PBHs or PBHs with massive UCMHs ($\delta m\gg M_{\rm PBH}$)
give $\eta f_{\rm PBH}/f_{\rm UCMH}\ll1$. Moreover, we
simply introduce a critical redshift $z_{m}$ below which ($z<z_{m}$)
X-ray emission from the super-Eddington accretion PBHs
($\dot{m}\gg1$) become important to heat and ionize the early Universe.
Hotter IGM heated by other energy sources (e.g., annihilation) slightly
decreases $z_{m}$. \cite{Ricotti08} showed that the
accretion becomes $\dot{m}\gg1$ at $z_{m}\sim20$ (100) for $M_{\rm
PBH}=10^{2}$ (300) $M_{\odot}$, and $\dot{m}$ is always
$\dot{m}\gg1$ ($\dot{m}\ll1$) for $M_{\rm PBH}>10^{3}M_{\odot}$
$(M_{\rm PBH}\ll10M_{\odot})$. However, as we discussed in Section
\ref{secAccPBH3}, the stage of super-Eddington accretion heating the IGM can be
delayed, because X-ray photons are trapped inside the total or inner region of the UCMHs due to
the accumulation and virialization of the accretion gas. Outflows can also (partly) absorb X-rays.
We take the critical redshift $z_{m}$ as showed in equations (\ref{gaccre06}) and (\ref{gaccre10}),
for $z<z_{m}$ X-ray photons from the super-Eddington accretion PBHs could escape their host
UCMHs. If the PBH abundance is much less than that of UCMH, but still satisfies
equation (\ref{PBH04}), X-ray emission should dominate over the early dark matter
annihilation at $z<z_{m}$.

\section{Reionization and Heating of the IGM}\label{secIonHeat}

\subsection{Basic Equations}\label{secIonHeat1}

The evolution of baryon ionization fraction $x_{\rm ion}(z)$ is given by
the differential equation (e.g., \citealt{CIP09})
\begin{equation}
n_{A}(1+z)^{3}\frac{dx_{\rm ion}(z)}{dt}=I(z)-R(z),\label{baseq01}
\end{equation}
where $I(z)$ and $R(z)$ are the ionization and recombination rates
per volume respectively, $n_{A}$ is the atomic number density today.
We use the rate $R(z)$ from \cite{NS08}.
The ionization rate per volume due to dark matter annihilation or
X-ray emission from gas accretion is given by
\begin{equation}
I(z)=\int_{E_{\rm eq}}^{E_{\chi}}dE_{\gamma}\frac{dn(z)}{dE_{\gamma}}P(E_{\gamma},z)N_{\rm ion}(E_{\gamma}),\label{baseq02}
\end{equation}
where $E_{\chi}=m_{\chi}c^{2}$ is the maximum energy of the emitted
photon, $E_{\rm eq}\simeq E_{\chi}(1+z)/(1+z_{\rm eq})$, the
differential term $dn(z)/dE_{\gamma}$ is the photon spectral number
density at redshift $z$. We follow \citealt{CIP09} (see also
\citealt{BH09}; \citealt{NS08,NS09}) to calculate the probability of
primary ionizations per second $P(E_{\gamma},z)$, and the number of
final ionizations that generated by a single photon of energy
$E_{\gamma}$ produces $N_{\rm ion}(E_{\gamma})$. Note that $N_{\rm
ion}(E_{\gamma})$ is proportional to the ionization factor
$\eta_{\rm ion}(x_{\rm ion})\approx(1-x_{\rm ion})/3$
(\citealt{SS85}; \citealt{Chen04}), which means approximately 1/3
emitted energy goes into the reionization of atoms if $x_{\rm
ion}\ll 1$.

First of all, we consider the ionization is due to UCMH dark matter
annihilation. The spectral number density is obtained as
\begin{equation}
\frac{dn_{\rm ann}(z)}{dE_{\gamma}}=\int_{\infty}^{z}dz'\frac{cdt}{dz'}\frac{dl_{\rm ann}(z')}{E_{\chi}dE_{\gamma'}(z')}\left(\frac{1+z}{1+z'}\right)^{3}
\exp(-\tau),\label{baseq03}
\end{equation}
where $l_{\rm ann}$ is the annihilation luminosity as mentioned in
Section \ref{secanni}. $E_{\gamma'}(z')=E_{\gamma}(1+z')/(1+z)$. The
optical depth $\tau$ is
\begin{equation}
\tau=\int_{z'}^{z}dz''\frac{cdt}{dz''}n_{A}(1+z'')^{3}\sigma_{\rm tot}(E_{\gamma}'')\label{baseq04}
\end{equation}
with $E_{\gamma''}=E_{\gamma'}(1+z'')/(1+z')$. The total cross
section $\sigma(\rm tot)$ for the DM annihilation photon to interact
with electrons in the IGM mainly includes the Klein-Nishina cross
section for Compton scattering (\citealt{Rybicki04}) and the
photonionization cross section for H and He as $\sigma_{\rm H+H_{\rm
e}}$ (\citealt{ZS89}). Pair production on matter becomes important
for $m_{\chi}>1$ GeV. CMB photons also contribute to the total cross
section for $m_{\chi}>10$ TeV, which can be neglected in our cases.

The total energy deposition per second per volume at redshift $z$ is
given by
\begin{equation}
\epsilon(z)=\int_{E_{\rm eq}}^{E_{\chi}}dE_{\gamma}\frac{dn_{\rm ann}(z)}{dE_{\gamma}}n_{A}(1+z)^{3}\sigma_{\rm tot}(E_{\gamma})E_{\gamma}\label{bkgd02}.
\end{equation}
If we take the monochromatic dark matter annihilation emission
for simplicity, i.e., the photons produced by dark matter
annihilation are the rest energy of the dark matter particle
$m_{\chi}c^{2}$, the photon flux spectral density then can be
calculated by the $\delta$-function
\begin{equation}
\frac{dl_{\rm ann}}{dE_{\gamma}'}(z')\approx l_{\rm ann}(z')\delta\left(E_{\gamma}'-E_{\chi}\right)\label{baseq05}.
\end{equation}
Thus we have the energy deposition $\epsilon(z)$ as
\begin{equation}
\epsilon(z)=\int_{E_{\rm eq}}^{E_{\chi}}\frac{dE_{\gamma}}{E_{\chi}}n_{A}(1+z)^{4}\frac{cdt}{dz_{0}'}l_{\rm ann}(z_{0}')\left(\frac{1+z}{1+z_{0}'}\right)^{3}
\textrm{e}^{-\tau}\sigma_{\rm tot}(E_{\gamma}),\label{emission}
\end{equation}
where $z_{0}'$ satisfies
\begin{equation}
z_{0}'=\frac{E_{\chi}}{E_{\gamma}}(1+z)-1\label{baseq06},
\end{equation}
and $\tau$ is calculated from $z_{0}'$ to $z$. Keep in mind in the
above formula (\ref{emission}) the dark matter annihilation products
are simplified as the gamma-ray photons with sole energy
$E_{\chi}=m_{\chi}c^{2}$. More realistic annihilation spectrum is
model-depended. For example, $dn/dE_{\gamma}$ can be chosen as
following the model in \cite{Bergstrom98} and \cite{Feng01}.

For the X-ray photons from an accreting PBH, equation (\ref{emission})
will still be available if we choose
$\sigma_{\rm tot}$ as the X-ray total cross section $\sigma_{\rm
tot}\approx\sigma_{\rm H+H_{\rm e}}+\sigma_{T}$, and $E_{\chi}$ with
the X-ray characteristic energy $E_{X}$ as $E_{X}\simeq3$ keV
$(M_{\rm PBH}/M_{\odot})^{-1/4}$ (\citealt{Salvaterra05}). As the
real accreting PBH spectral energy distribution is very
model-depended (e.g., \citealt{SS73}; \citealt{Sazonov04};
\citealt{Salvaterra05}; \citealt{RMF08}), in the very first
calculation we simplify the X-ray emission as the single-frequency
emission at a characteristic energy $E_{X}$, which mostly can be
considered as the peaked energy in the real spectral energy
distribution. We will also discuss the more realistic PBH spectral
energy distribution in the discussion section \ref{SecDis03}.

Now we list the heating and cooling processes in the IGM. The
heating of IGM by UCMH annihilation or X-ray emission can be written
as
\begin{equation}
\left(\frac{dT_{\rm m}}{dt}\right)_{\rm ann}=\frac{2}{3k_{B}}\frac{\eta_{\rm heat}(x_{\rm ion})}{n_{A}(1+z)^{3}(1+f_{\rm He}+x_{\rm ion})}\epsilon(z),\label{heat01}
\end{equation}
where the heating fraction $\eta_{\rm heat}(x_{\rm ion})$ which
shows the portion of energy $\epsilon(z)$ into heating IGM is
adopted as $\eta_{\rm heat}=C(1-(1-x_{\rm ion}^{a}))^{b}$ with
$C=0.9971$, $a=0.2663$ and $b=1.3163$ (\citealt{SS85}). We can
approximated take the He fraction in the IGM as $f_{\rm
He}\simeq0.073$. Moreover, CMB photons can be treated as another
heating source for the IGM if the IGM gas is colder than the CMB
($T_{\rm m}<T_{\rm CMB}$), otherwise the IGM gas would transfer
energy into the CMB environment. The coupling between IGM gas and
the CMB photons can be important when the difference between $T_{\rm
m}$ and $T_{\rm CMB}$ is significant (\citealt{Weynmann65};
\citealt{Tegmark97}; \citealt{SSS00})
\begin{equation}
\left(\frac{dT_{\rm m}}{dt}\right)_{\rm comp}\approx k_{\rm comp}T_{\rm CMB}^{4}x_{\rm ion}(T_{\rm CMB}-T_{\rm m})\label{heat02},
\end{equation}
where the coupling rate coefficient $k_{\rm comp}\simeq
5.0\times10^{-22}$ s$^{-1}$.

Other IGM cooling terms are dominated by the adiabatic cooling
during the expansion of the universe as
\begin{equation}
\left(\frac{dT_{\rm m}}{dz}\right)_{\rm ad}=\frac{2T_{\rm m}}{1+z},
\end{equation}
for low temperature. Note that the IGM temperature will decrease
independently as $T_{\rm m}\propto(1+z)^{2}$ for a pure adiabatic
cooling process. Furthermore, for sufficient high temperature
$\sim10^{4}$ K, the molecular hydrogen H$_{2}$ cooling will also be
important. This cooling term can be calculated as
\begin{equation}
\left(\frac{dT_{\rm m}}{dt}\right)_{\rm H_{2}}=\Lambda_{\rm H_{2}}(1-x_{\rm ion}-2f_{\rm H_{2}})f_{\rm H_{2}}[n_{A}(1+z)^{3}]^{2}\label{heat03},
\end{equation}
where we adopt the specific cooling coefficient $\Lambda_{\rm
H_{2}}$ from \cite{Hollenback79} and \cite{Yoshida06}. We neglect
other chemical cooling processes such as Bremsstrahlung, helium line
cooling, H$_{2}$ line cooling and hydrogen three-body reaction. HD
cooling is important for $T<200$ K and low density
(\citealt{Yoshida06}), but the gas adiabatic cooling will be
dominated in this case. As we mainly pay attention to the evolution
of the ionized fraction $x_{\rm ion}$ and IGM temperature $T_{\rm
m}$, we only include the evolution of hydrogen (H, H$^{-}$, H$^{+}$,
H$_{2}$) and electron gas ($e^{-}$) as the main species for
ionization. More detailed simulation including other species and
cooling processes is beyond our purpose of this paper. The evolution
of the H$_{2}$ fraction is adopted from the semi-analytic model in
\cite{Tegmark97}
\begin{equation}
\frac{df_{\rm H_{2}}}{dt}=k_{m}n_{A}(1+z)^{3}(1-x_{\rm ion}-2f_{\rm H_{2}})x_{\rm ion}\label{heat04},
\end{equation}
where we follow \cite{Tegmark97} and \cite{Galli98} to calculate the
reaction coefficient $k_{m}$.

\subsection{Solutions of IGM Evolution}\label{secsol}

\begin{figure}
\centerline{\includegraphics[width=100mm]{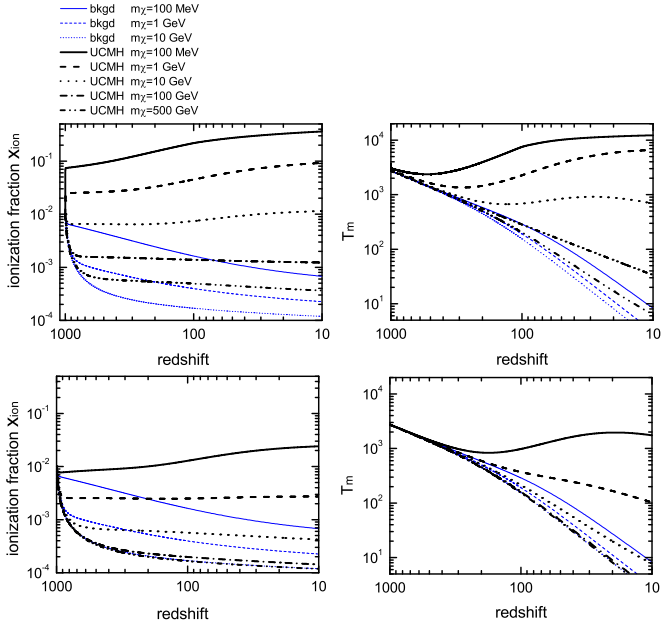}}
\caption{Effects of UCMH dark matter annihilation on the IGM evolution. Upper two panels correspond
to the case of $f_{\rm UCMH}=10^{-4}$ and lower to $f_{\rm UCMH}=10^{-6}$. The thick (black) lines
from the top down show the results
for UCMH annihilation with $m_{\chi}c^{2}$=100 MeV, 1 GeV, 10 GeV, 100 GeV and 500 GeV, while the thin (blue)
lines give the results of homogenous dark matter background annihilation background
for $m_{\chi}c^{2}$=100 MeV, 1 GeV and 10 GeV.}\label{figIGMhalo}
\end{figure}

Since no direct evidence related to the UCMH radiation has been confirmed
until now, the UCMH abundance is still uncertain. In this paper
we take the UCMH fraction $f_{\rm UCMH}$ as a free parameter. In
Fig. \ref{figIGMhalo} we show the ionization fraction $x_{\rm
ion}(z)$ and the IGM temperature $T_{\rm m}$ for $f_{\rm
UCMH}=10^{-4}$ and 10$^{-6}$, which correspond to today's expected
abundance $\sim1\%$ and $10^{-4}$ respectively. Moreover general discussion
on the UCMH abundance will be given later. The initial $x_{\rm ion}$ at $z=1000$ is adopted as
0.01, and $T_{\rm m}$ as the CMB temperature (\citealt{Galli98};
\citealt{Ripamonti07}; \citealt{RMF07a,RMF07b}). The first basic
conclusion which is similar to the previous works is that, lighter
dark matter particles or higher UCMH abundance give larger
$x_{\rm ion}$ and higher $T_{\rm m}$. An extreme bright UCMH
annihilation background (e.g., $m_{\chi}c^{2}\leq1$ GeV for $f_{\rm
UCMH}\simeq10^{-4}$ or $m_{\chi}c^{2}\leq100$ MeV for $f_{\rm
UCMH}\simeq10^{-6}$), even gives a monotonically increased $x_{\rm
ion}$ and $T_{\rm m}\geq10^{4}$ K without a standard reionization
epoch in the early Universe. We compare the UCMHs annihilation results
with the homogenous background annihilation, note that the homogenous dark
matter annihilation background only produce noticeable effects for
light dark matter particles $m_{\chi}c^{2}<1$ GeV or sterile
neutrinos. UCMHs, which provide a new dominated dark matter
annihilation gamma-ray background as showed in Section
\ref{secanni}, play a more important role to ionize and heat the early Universe.

\begin{figure}
\centerline{\includegraphics[width=100mm]{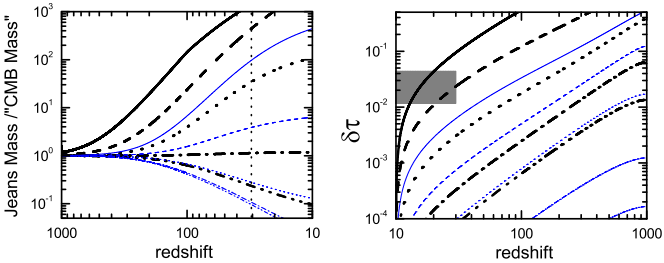}}
\caption{Left: ratio of Jeans mass to the ``CMB mass'', which is the Jeans mass for the gas temperature always being
equal to the CMB temperature. Right: evolution of the remaining Thomson scattering optical depth $\delta\tau$
with gray belt showing the WMAP 5-year  $1\sigma$ of remaining CMB optical depth at $10\leq z<30$.
The thick (black) lines are for $f_{\rm UCMH}=10^{-4}$ and thin (blue) for $f_{\rm UCMH}=10^{-6}$.
The lines with same color from the top down are for $m_{\chi}c^{2}=100$ MeV to 500 GeV as in Fig. \ref{figIGMhalo}.}\label{figHaloJM}
\end{figure}

Furthermore, as the cosmological Jeans mass $m_{J}$ can be taken as an indicator of
the IGM structure evolution, the left panel of Fig.
\ref{figHaloJM} gives the evolution of Jeans mass in the cosmic
UCMH annihilation background. The Jeans mass $m_{J}\propto
T_{m}^{3/2}\rho^{-1/2}$ should be a constant if the gas temperature
is always equal to the CMB temperature $T_{\rm m}=T_{\rm CMB}$. In
this paper we call this constant as ``CMB mass''. In the left panel
of Fig. \ref{figHaloJM}, $m_{J}$ with various
$T_{\rm m}$ is generally normalized in the unit of ``CMB mass''. The
remaining Thomson scattering optical depth contributed by UCMH
annihilation is showed in the right panel of Fig. \ref{figHaloJM}.
For $6\leq z<30$, the remaining CMB optical depth is estimated as
$\delta\tau\simeq0.046\pm0.016$ by the WMAP five-year
measurement.\footnote{The WMAP five-year measurements give the CMB
Thomson scattering optical depth $\tau\simeq0.084\pm0.016$
(\citealt{Komatsu09}), which is mostly due to ionization at late
times $z<30$ (\citealt{Ricotti08}; \citealt{NS10}). If we subtract
the optical depth contributed by the totally ionized gas
$\tau(z\leq6)=0.038$, the remaining depth is
$\delta\tau\simeq0.046\pm0.016\leq0.062$ (\citealt{CIP09}).} We
assume a linear increase of $x_{\rm ion}$ from $z=10$ to the full
ionization time $z=6$, thus the upper bound contribution of
UCMH annihilation to the measurable CMB optical depth is
$\delta\tau\simeq0.028\pm0.016\leq0.044$. Based on this consideration, in our examples
only one extreme case that $m_{\chi}c^{2}=100$ MeV with $f_{\rm
UCMH}=10^{-4}$ is ruled out by the CMB remaining optical depth in
Fig. \ref{figHaloJM}. UCMH
annihilation can significantly increase the Thomson optical depth in the early Universe $z\gg100$
up to $\delta \tau\sim0.5$ without stringent
constraints by the CMB optical depth measurement at $z<30$. After the last-scattering epoch the ionization
fraction $x_{\rm ion}$ can change from $x_{\rm ion}\sim10^{-4}$
(without dark matter annihilation) to a upper bound $x_{\rm
ion}\sim0.1$ (e.g, $m_{\chi}c^{2}=1$ GeV and $f_{\rm
UCMH}=10^{-4}$). Also, the IGM can be heated from a temperature
$T_{\rm m}$ of adiabatically cooling $T_{\rm m}\propto(1+z)^{2}$ in
the absence of heating source to the upper bound $T_{\rm
m}>10^{3}$ K with a sufficient amount of heating contributed by UCMH
annihilation. An much higher Jeans mass than the ``CMB mass'' by $\sim2-3$
orders of magnitude increase
can be obtained due to the hotter IGM temperature. Therefore, we can natively
estimate that the formation of small baryonic objects can be
strongly suppressed, although more investigations need to be carried
out in Section \ref{secStruc}.

In general, we find the impact of UCMH annihilation on the IGM
evolution can be empirically estimated by the factor $m_{\chi,100}^{-1}f_{\rm UCMH}$,
while the threshold of UCMH abundance to affect the IGM evolution is
approximately given by $m_{\chi,100}^{-1}f_{\rm UCMH}>10^{-6}$, with an
upper bound constrained by the CMB optical depth at late times $z<30$ as $x_{\rm
ion}\sim0.1$ and $T_{\rm m}\sim5000$ K at $m_{\chi,100}^{-1}f_{\rm UCMH}\sim10^{-2}$. The CMB optical depth
enhancement at early times $z>30$ can be more dramatic than the late
times due to the higher annihilation luminosity in early redshift
(equation {\ref{ann01}}), which is different from the PBH radiation
which has higher luminosity at late times (Section \ref{secAccPBH3}; \citealt{Ricotti08}). Further phenomenological
constraints should be made by CMB polarization anisotropies, which
is left for a future investigation. Keep in mind another cannel to
concentrate dark matter rather than primordial density perturbation
is the formation of the first dark objects, which should affect the
IGM evolution much later ($z<100$) than UCMHs. Therefore UCMH
annihilation has a definitely much earlier and more important impact
on the IGM evolution from the last scattering to the structure formation time.

So far we give the results only for UCMH dark matter annihilation.
Whether the X-ray emission from the PBH host UCMHs will
significantly change the above results mainly depends on the
fraction of PBH $f_{\rm PBH}$, the average inflow radiation efficiency $\eta$ and the critical redshift $z_{m}$ as
given in Section \ref{secAccPBH}. In our paper we combine the factor
$\eta f_{\rm PBH}$ as one. Remember the
results in Section \ref{secAccPBH}: when X-rays from the center PBH
region successfully passes through the transparent baryon medium in
the host UCMH at $z<z_{m}$, the much brighter X-ray luminosity and
much larger interacting cross section $\sigma_{\rm tot}(E_{X})$
compared to the annihilation luminosity and $\sigma_{\rm
tot}(E_{\gamma})$ usually guarantees the X-ray emission to be
dominated over the UCMH annihilation (equations \ref{PBH03}), except for a much lower PBH
fraction $f_{\rm PBH}$ below the value in equation (\ref{PBH04}).
We will give a lower limit of $f_{\rm PBH}$, above which X-rays have
obvious impact on the IGM evolution at $z<z_{m}$. In the following calculation we
assume equation (\ref{PBH04}) is always satisfied, and do not
distinguish the redshift which divide the UCMH radiation into
annihilation dominated or X-ray radiation dominated from the redshift which gives a
transparent baryon environment in UCMHs, but simply use one parameter $z_{m}$.

\begin{figure}
\centerline{\includegraphics[width=100mm]{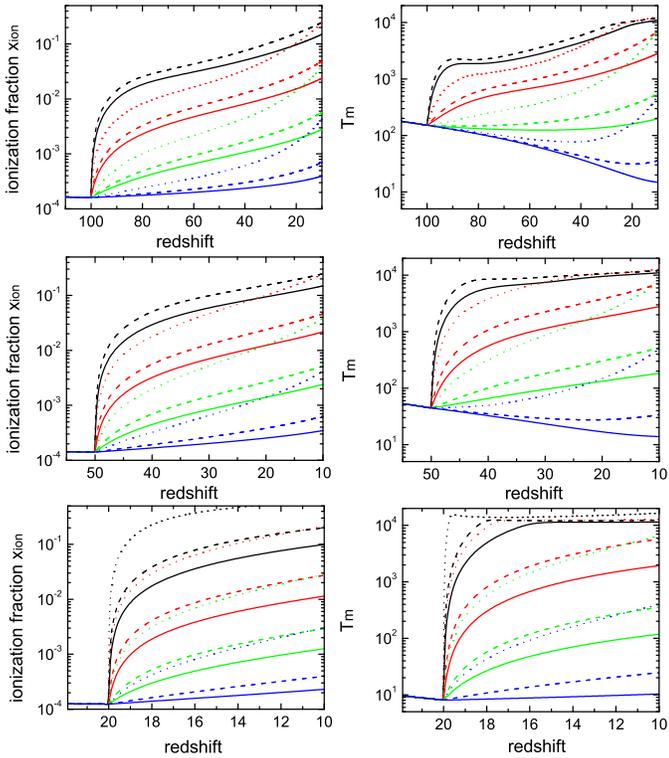}}
\caption{Effects of gas accretion around PBHs on the IGM ionization and temperature evolution.
The upper panels are for the X-ray energy injection starting at $z_{m}=100$, middle panels for
$z_{m}=50$ and lower for $z_{m}=20$. The PBH fraction $\eta f_{\rm PBH}=10^{-7}$ (black lines),
$10^{-8}$ (red lines), $10^{-9}$ (green lines) and $10^{-10}$ (blue lines). Characteristic emission
energy are 100 keV (solid lines), 10 keV (dashed lines) and 1 keV (dotted lines).}\label{figIGMPBH}
\end{figure}

\begin{figure}
\centerline{\includegraphics[width=100mm]{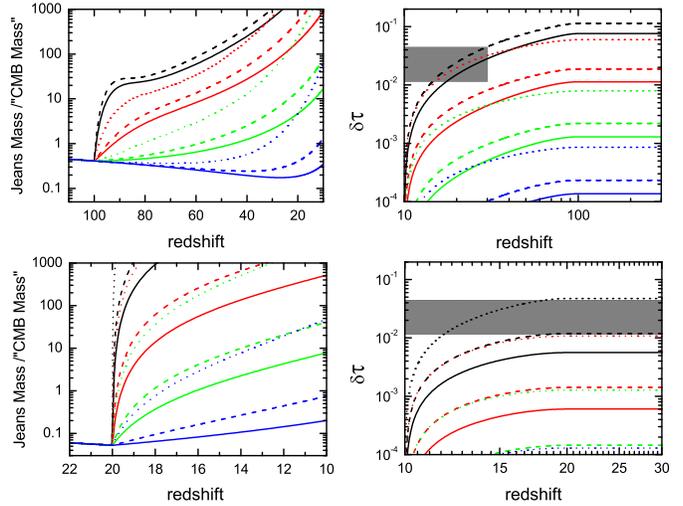}}
\caption{Ratio of Jeans mass to the ``CMB mass'' (left panels) and evolution
of Thomson scattering depth (right panel) for the starting injection redshift
$z_{m}=100$ (upper panels) and 20 (low panels). The lines are as the same in Fig. \ref{figIGMPBH}.}\label{figPBHJM}
\end{figure}

In Fig. \ref{figIGMPBH} we take $z_{m}$ and $\eta f_{\rm PBH}$ as
parameters with the characteristic emission frequency as $E_{X}=1$
keV, 10 keV and 100 keV, which correspond to the typical PBH mass as
$10^{2}M_{\odot}$, $10^{-2}M_{\odot}$ and $10^{-6}M_{\odot}$
respectively. Higher $z_{m}$ means higher ratio $\delta m/M_{\rm PBH}$
or lighter dark matter particles. Only X-ray emission as the energy source is calculated
in this figure \footnote{The more realistic case is that both $z_{m}$ and $E_{X}$ are the functions of the
mass of PBH and its host UCMH, thus in principle neither $z_{\rm m}$ nor
$E_{X}$ are single free parameters.
The real X-ray radiation background due to the PBH accretion happens
at a certain $z_{m}$ as showed in equations (\ref{gaccre06}) and
(\ref{gaccre10}), and change its spectral energy distribution as a
function of redshift (\citealt{RMF08}). Technically it is able to calculate the X-ray
emission variation by introducing the initial mass function of PBHs and the host UCMHs.
However, both of these two initial mass functions are poorly known.
More elaborate models only make our model more
complicated and uncertain. What we focus on in our calculations is
the key differences between X-ray radiation and the much earlier
occurred UCMH annihilation, so we only use
the characteristic energy $E_{X}$, and the ratio $\eta f_{\rm PBH}/f_{\rm UCMH}$ to
show the importance of the X-ray radiation.
More discussion of PBH spectral energy distribution can be seen in Section
\ref{SecDis03}.}. As showed in Fig. \ref{figIGMPBH}, the final
properties of the IGM at $z\sim10$ with same $\eta f_{\rm PBH}$ and
$E_{X}$ are more or less closed to each other regardless the value
of $z_{m}$, which means lower $z_{m}$ gives more dramatic thermal
and chemical change at $z<z_{m}$. According to equation (\ref{gaccre10}), lower $z_{m}$
corresponds to lighter $m_{\chi}$, which also increase the UCMH annihilation.
On the other hand, $x_{\rm ion}$ and $T_{\rm m}$ vary for
more than three orders of magnitude from $E_{\rm X}\sim100$ keV
($10^{-6}M_{\odot}$) to $E_{\rm X}\sim1$ keV ($10^{2}M_{\odot}$)
with the same $\eta f_{\rm PBH}$, which means massive PBH favors the IGM ionization.
On the other hand, the X-ray radiation effect can
be neglected when $\eta f_{\rm PBH}\leq10^{-11}$ for $E_{X}\sim100$
keV, but a smaller limit $\eta f_{\rm PBH}\leq10^{-12}$ is
applied for $E_{X}\sim1$ keV. Below the lower limit the PBHs are not
expected to have any promising effects on reionization. The estimate
in Fig. \ref{figPBHJM} shows that no strict constraints are made for
$\eta f_{\rm UCMH}\leq10^{-7}$ by the remaining CMB depth $\delta
\tau$, which allows a dramatically increased Jeans mass due to the
hot IGM gas $\sim10^{4}$ K.

As a result, X-ray emission from PBHs gives a more
promising impact on the IGM evolution if $\eta f_{\rm
PBH}\gg10^{-11}$ ($10^{-12}$) for $M_{\rm PBH}\sim10^{-6}M_{\odot}$
($10^{2}M_{\odot}$), or empirically to say, $\eta f_{\rm
PBH}\gg1.8\times10^{-12}(M/M_{\odot})^{-1/8}$. As we assume $f_{\rm UCMH}\leq10^{-4}$, we
expect the UCMH dark matter annihilation only played its role on the
IGM evolution at very high redshift $z_{m}<z<1000$, but X-ray
emission changes the Universe reionization history dramatically at
relatively lower redshift $z<z_{m}$. Considering $\eta\sim0.1$, the upper bound
value $f_{\rm PBH}\leq10^{-6}$ is two orders of magnitude higher than
the upper PBH abundance $\leq10^{-8}$ in \cite{Ricotti08} for
$M_{\rm PBH}>10^{3}M_{\odot}$, but much lower than the low mass PBH abundance
constraint in \cite{Ricotti08}. However, the massive PBH abundance constraint in \cite{Ricotti08}
is made by the Compton y-parameter estimate at $z_{\rm rec}<z<z_{\rm eq}$ without local
UCMH trapping, while we consider the two-step accretion first by the host UCMHs and
then by the center PBHs as mentioned in Section \ref{secAccPBH3},
X-rays can just locally heat the accreted gas inside UCMHs but not the entire cosmic gas at high redshift.

\begin{figure}
\centerline{\includegraphics[width=100mm]{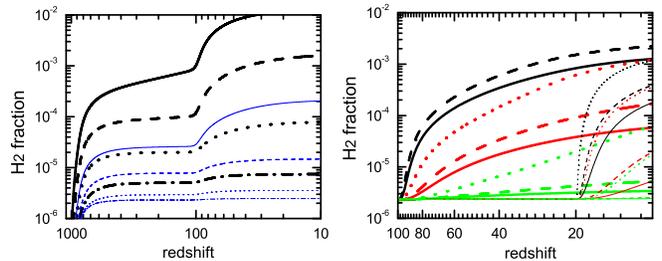}}
\caption{H$_{2}$ abundance evolution for UCMH annihilation (left panel) and gas accretion
onto PBHs (right panel). The lines in the left are the same as in Fig \ref{figHaloJM}.
The lines in the right panel are the same as in Fig.
\ref{figIGMPBH}, but thick and thin are for $z_{m}=100$ and $z_{m}=20$ respectively.}\label{figIGMH2}
\end{figure}

The abundance of IGM molecular hydrogen $f_{\rm H_{2}}$ in various
UCMH radiation models is showed in Fig. \ref{figIGMH2} in this
section. We see that when the UCMH energy injection can be
neglected, this fraction goes back to $f_{\rm H_{2}}\sim10^{-6}$,
which is consistent with the standard result (e.g.,
\citealt{Galli98} and references therein). The upper bound of enhanced $f_{\rm H_{2}}$
is $f_{\rm H_{2}}\sim10^{-3}$, either due to the allowed UCMH dark matter annihilation
constrained by the CMB optical depth, or the X-ray emission at $z_{m}\leq100$.

\section{First Structures}\label{secStruc}

In the hierarchical cold dark matter (CDM) scenario, the first
cosmological objects are dark matter haloes, which are formed by
gravitational instability from the scale-free density fluctuations
(e.g., \citealt{Green05}; \citealt{Diemand05}; \citealt{Yoshida09}).
The formation of the first baryonic objects depends on the detailed
gas dynamical processes. The first baryonic objects can successfully
collapse and form inside dark haloes when its cooling timescale for dissipating
the kinetic energy is much shorter than the Hubble time.
\cite{Tegmark97} showed that the formation of the baryonic structure
crucially depends on the abundance of the molecular hydrogen
$f_{\rm{H}_{2}}$. \cite{Biermann06} considered that the photons
emitted by dark matter annihilation or decay inside the halo can
boost the production of H$_{2}$, and may favor the formation of the
first structure. On the other hand, a different conclusion that the
dark matter annihilation or decay can slightly delay the first
baryonic structure formation was given by \cite{RMF07b}. As they
discussed, the higher central density in the baryonic cloud without
dark matter energy injection could compensate the lower abundance of
H$_{2}$ and still lead to the fastest cooling. Nevertheless, no matter
promotion or suppression caused by halo extended dark matter annihilation or
decay, such effects are pretty small.

UCMHs can also be captured by the first dark matter
objects, in this case the UCMH radiation can be much brighter than that
from the extended large dark matter halo, even the fraction of
UCMHs in the dark halo is tiny. In this section we fucus on the dark halo
structures with mass $\sim10^{6}M_{\odot}$ (or $10^{7}M_{\odot}$ in the PBH heating case),
as they favor the later first star formation (\citealt{Bromm09}; \citealt{Yoshida09}). For a
typical $10^{6}M_{\odot}$ halo in the early Universe,
$\sim10^{20}$ ($10^{18}$) or $\sim10^{8}$ ($10^{6}$) UCMHs within
this halo can be expected for an initial UCMH fraction $f_{\rm
UCMH}\sim10^{-4}$ ($10^{-6}$) if the seed of UCMHs are generated in
the electroweak or QCD phase transitions respectively
(\citealt{Scott09}). In these cases the UCMH emission can also
provide a new type of radiation background in the dark halo.
However, the uniform UCMH distribution treatment will break down if the number of
massive UCMHs in a halo is less than $\sim10$. This happens for the massive UCMH case
that $f_{\rm UCMH}M_{\rm DM}/m_{h}\leq10$. In this section we study the effects of UCMH radiation on the first
structure formation and evolution with $f_{\rm UCMH}M_{\rm
DM}\gg m_{h}$, the case of only-several-luminous-UCMH will be discussed separately
in Section \ref{SecDis04} as a supplementary. As the virial temperature of
$\sim10^{6}M_{\odot}$ haloes is less than the threshold for
atomic hydrogen line cooling, these haloes are often referred as
``minihalo'' in literatures. However, for clearly, in this paper we
call the first dark matter structure as (cosmological) dark matter haloes or
dark haloes, which should not be confused with the UCMHs.

\subsection{Dark Matter Annihilation}\label{secStruc01}

\begin{figure}
\centerline{\includegraphics[width=90mm]{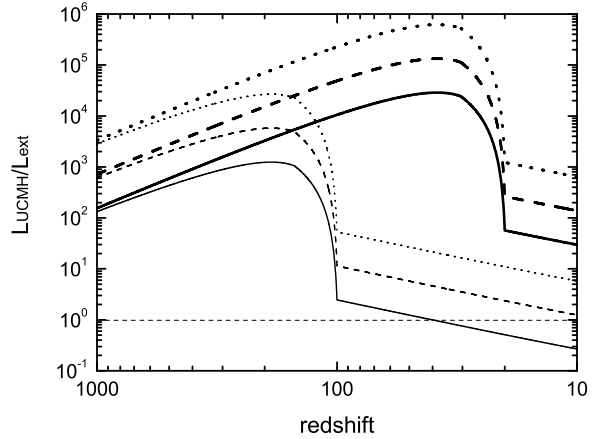}}
\caption{Ratio between $L_{\rm UCMH}$ and $L_{\rm ext}$ for
$m_{\chi}c^{2}=$1 GeV (solid lines), 10 GeV (dashed lines), 100 GeV (dotted lines) and isothermal extended dark matter halo.
Thick and thin lines correspond to the virial redshift $z_{\rm vir}=20$ and 100 respectively.
The UCMH fraction is adopted as $f_{\rm UCMH}=10^{-6}$. The case of $L_{\rm UCMH}=L_{\rm ext}$ is
marked as the horizon thin dashed line.}\label{fighaloratio}
\end{figure}

The profiles of the cosmological dark matter haloes are chosen
before our calculation. The equations for the halo profile are
listed in the Appendix. Before the formation of the first stars, the
energy injection inside a large dark matter halo mainly contributed
by the local emission from the UCMHs in the halo, the local
annihilation or decay of the extended dark matter within
the halo, and the outside radiation background which injects into
the halo. We check the total energy produced by dark matter
annihilation with a dark halo as $L_{\rm halo}=L_{\rm UCMH}+L_{\rm
ext}$, with $L_{\rm UCMH}$ and $L_{\rm ext}$ being the annihilation
luminosity from UCMHs and the extended dark matter in this halo. The
typical ratio $L_{\rm UCMH}/L_{\rm ext}$ is demonstrated in Fig.
\ref{fighaloratio}, where we adopt an isothermal dark halo model and
$f_{\rm UCMH}=10^{-6}$. We find both $L_{\rm UCMH}$ and $L_{\rm ext}$ are
proportional to the total halo mass $M_{\rm DM}$, so $L_{\rm
UCMH}/L_{\rm ext}$ is independent to $M_{\rm DM}$. In
Fig. \ref{fighaloratio} $L_{\rm UCMH}$ is mostly dominated
in the halo, except for large $z_{\rm vir}$ with light dark matter particles(e.g.,
$<10$ GeV for $f_{\rm UCMH}=10^{-6}$ and $z_{\rm vir}=100$, much
lighter dark matter particles is required for more abundant UCMHs or
smaller $z_{\rm vir}$). Therefore, similar to the IGM environment,
the energy injection mechanism inside a dark halo which contains
UCMHs can be very different from the no-UCMH case. We focus on the
ionization and heating inside the dark matter halo. Also, we mention
that the results for $L_{\rm UCMH}/L_{\rm ext}$ in NFW haloes are
very similar to the isothermal haloes in Fig. \ref{fighaloratio}.
\begin{figure}
\centerline{\includegraphics[width=90mm]{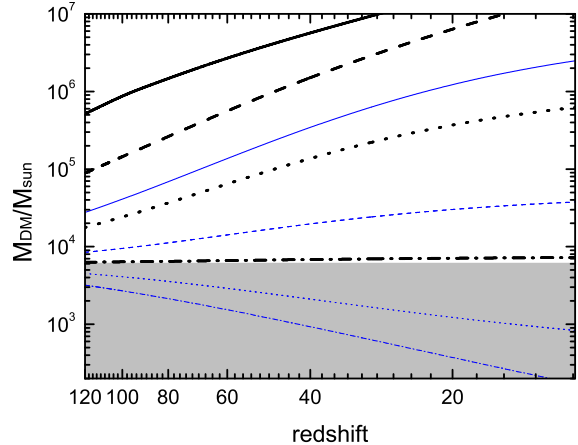}}
\caption{The minimal dark halo mass for $T_{\rm m}=T_{\rm vir}$ with the IGM heated by UCMHs.
The shaded area is for the case that the minimal halo mass for gas collapsing becomes
$M_{\rm DM}<6.1\times10^{3}M_{\odot}$ due to a lower ambient gas temperature
compared to the CMB temperature. The lines from $m_{\chi}c^{2}=100$ MeV to 500 GeV as in Fig. \ref{figHaloJM}.}\label{figMhalo_A}
\end{figure}

The simple criterion of baryonic matter filling in a dark halo
is approximately taken as the IGM gas temperature being cooler than
the virial temperature of the halo $T_{\rm m}<T_{\rm vir}$,
otherwise the gas pressure prevents the gas from collapsing. We do not include the
temperature cooling slope criterion as in \cite{Tegmark97} to
further study the baryon cooling and collapsing in the dark halo. Fig.
\ref{figMhalo_A} gives the dark matter halo mass for the critical
case $T_{\rm m}=T_{\rm vir}$ with UCMH annihilation heating the
ambient IGM gas. Generally more massive dark halo mass is needed for
gas filling into dark haloes and forming baryonic structure in the
haloes. If $T_{\rm m}=T_{\rm CMB}$ the minimum halo mass for gas
filling in is $6.1\times10^{3}M_{\odot}$, while the minimum halo
mass increases significantly for $T_{\rm m}\gg T_{\rm CMB}$. In this
sense, the formation of the first baryonic objects will be obviously
suppressed in the small dark haloes located in host IGM gas. However, in
the pure UCMH annihilation case without PBH radiation, we expect
$\sim10^{6}M_{\odot}$ dark haloes attract baryons at $z>10$ in most cases
expect for $f_{\rm UCMH}m_{\chi,100}^{-1}\geq10^{-2}$.

The annihilation energy deposited in a dark halo can be linearly
divided into two parts: the energy from the background where the
cosmic UCMH annihilation occurs $\epsilon_{\rm bgd}(z)$, and that
within the local dark halo $\epsilon_{\rm loc}(z)$.
The term $\epsilon_{\rm bgd}(z)$ is obtained by equation
(\ref{bkgd02}). The local energy $\epsilon_{\rm loc}(z)$ is a
function of position inside the halo. We focus on the energy
deposition at the center of the halo. The contribution by the
extended dark matter in an isothermal halo is
\begin{equation}
\epsilon_{\rm loc,iso}(z)=\frac{\langle\sigma v\rangle c^{2}(1-f_{\chi})}{2\mu m_{p}m_{\chi}f_{\chi}}\sigma_{\rm tot}(E_{\chi})\rho_{\rm core}^{3}
\left(\frac{6}{5}R_{\rm core}-\frac{R_{\rm core}^{6}}{5R_{\rm tr}^{5}}\right),\label{structure01}
\end{equation}
where $f_{\chi}=\Omega_{\rm DM}/\Omega_{\rm M}\approx0.833$. On the
other hand, the local energy deposited by the UCMH annihilation
depends on the UCMH distribution inside the halo. If we assume the
UCMHs number density is uniformly distributed depending on the halo mass density, i.e.,
$\frac{dn_{\rm UCMH}(r)}{dM_{\rm DM}(r)}\propto const.$ (a relevant distribution simulation see
\citealt{Sandick11}), we have
\begin{eqnarray}
&&\epsilon_{\rm loc,UCMH}(z)\simeq\frac{L_{\rm UCMH}\sigma_{\rm tot}(E_{\chi})(1-f_{\chi})}{\mu m_{p}f_{\chi}M_{\rm DM}}\int_{0}^{R_{\rm tr}}\rho^{2}dr\nonumber\\
&&\simeq\frac{L_{\rm UCMH}\sigma_{\rm tot}\rho_{\rm core}^{2}(1-f_{\chi})}{\mu m_{p}M_{\rm DM}f_{\chi}}\left(\frac{4}{3}R_{\rm core}-\frac{R_{\rm core}^{4}}{3R_{\rm tr}^{3}}\right)
\label{deposition2}.
\end{eqnarray}
In the following calculation we adopt equation (\ref{deposition2}) for UCMH
annihilation, and also include the extended dark matter annihilation
within the halo.

\begin{figure}
\centerline{\includegraphics[width=100mm]{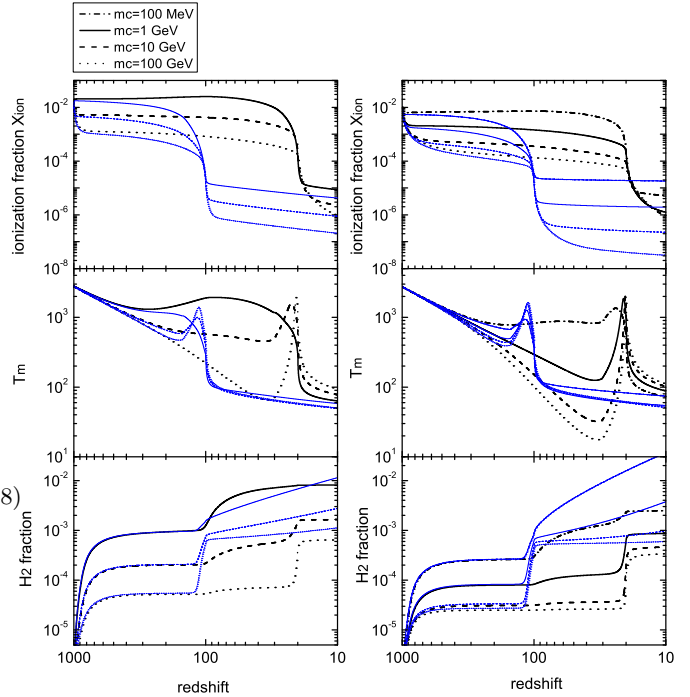}}
\caption{Effects of UCMH and extended dark matter annihilation inside an $10^{6}
M_{\odot}$ isothermal halo on the evolution of ionization (upper
panels), temperature (middle panels) and H$_{2}$ fraction $f_{\rm
H_{2}}$ (lower panels) at the central region of the halo,
where we choose the virial redshift $z_{\rm vir}=20$ (thick black
lines) and 100 (thin blue lines), $m_{\chi}c^{2}=100$ MeV (dash-dotted lines),
1 GeV (solid lines), 10 GeV (dashed lines) and 100 GeV (dotted lines).}\label{figShalo}
\end{figure}

Figure \ref{figShalo} shows the gas evolution at the center
of an $10^{6}M_{\odot}$ isothermal halo virializing at $z_{\rm
vir}=20$ or $z_{\rm vir}=100$. We also show the protohalo stage at
$z>z_{\rm vir}$. More energetic annihilation due to larger $f_{\rm UCMH}$
or lower $m_{\chi}$ gives higher $x_{\rm ion}$ and $f_{\rm H_{2}}$. The UCMH
annihilation gives a significant impact on $T_{\rm
m}$ before virialization in the protohalo stage, that is because the
cooling and heating mechanisms are different before and after
virialization. The change of $T_{\rm m}$ before virialization is
mainly due the heating by background UCMH annihilation, which
is totally dominated over the extended dark matter annihilation, thus
brighter UCMH annihilation luminosity gives a higher gas temperature
at $z>z_{\rm vir}$. However, after a dramatic temperature increase
during the virializing $z\sim z_{\rm vir}$, H$_{2}$ cooling becomes
the main process to cool the denser gas at $z<z_{\rm vir}$.
The peak temperature during virializing is around
$\sim1000-2000$ K. We find that higher H$_{2}$
abundance, which is caused by brighter UCMH annihilation, gives
a lower gas temperature after virialization for small $z_{\rm vir}$
($\sim20$), but a higher temperature for large $z_{\rm vir}$
($\sim100$). This result is just between that in \cite{Biermann06},
who considered the effects of sterile neutrino decay can favor the
structure formation, and \cite{RMF07b}, who showed that dark matter
annihilation will slight delay the structure formation. The main
reason of our difference from \cite{RMF07b} for $z_{\rm vir}\ll100$
is that, we take the baryon gas density to be proportional to
the halo density $n_{b}\propto\rho$ as in \cite{Tegmark97},
therefore more molecular gas due to stronger heating just means more
efficient cooling. A more elaborate result can be made by
adding more detailed gas dynamics and energy transfer including
the UCMH radiation within the halo (\citealt{Tegmark97};
\citealt{RMF07b}). But such a new calculation should not change the fact that
UCMH radiation, as well as the extended dark halo annihilation, cannot
change the gas temperature in the halo obviously after virialization.

Note that the temperature $T_{\rm m}$ in the halo only change by
a factor of $\sim3$ for a several orders of magnitude change to the UCMH
annihilation luminosity inside the halo. Therefore, we cannot expect the first baryonic
structure formation can be obviously promoted or suppressed. After
virialization, the effects of dark matter annihilation are always
secondary compared to the H$_{2}$ cooling mechanism. On the other
hand, gas chemical properties such as $x_{\rm ion}$ and $f_{\rm
H_{2}}$ can be changed significantly that higher $x_{\rm ion}$ and $f_{\rm{H}_{2}}$ are produced by brighter
dark matter annihilation.

\subsection{Gas accretion onto PBHs and X-ray Emission}\label{secStruc02}

\begin{figure}
\centerline{\includegraphics[width=90mm]{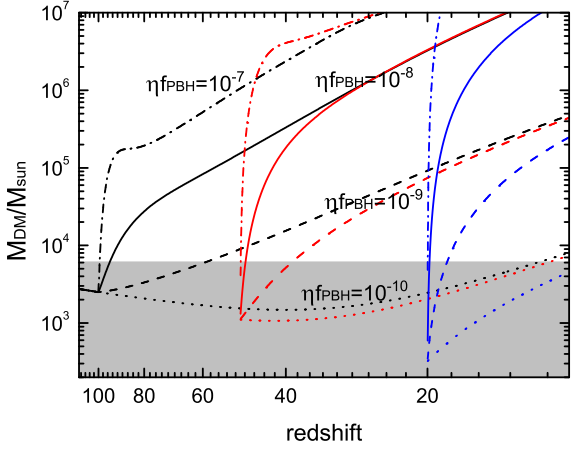}}
\caption{The minimal dark halo mass at $T_{\rm m}=T_{\rm vir}$ for PBH radiation with
$\eta f_{\rm PBH}=10^{-7}$ (dash-dotted lines), $10^{-8}$ (solid lines), $10^{-9}$ (dashed lines),
$10^{-10}$ (dotted lines) and $z_{m}=100$ (dark lines),
50 (red lines) and 20 (blue lines). The characteristic PBH radiation is $E_{X}=10$ keV.
The shaded area is for the case that the minimal halo mass for gas collapsing becomes
$M_{\rm DM}<6.1\times10^{3}M_{\odot}$ due to a lower ambient gas temperature
compared to the CMB temperature.}\label{figMhalo_B}
\end{figure}

If we consider gas accretion onto PBHs, and take the
PBH fraction as $\eta f_{\rm PBH}>10^{-11}$ $(10^{-12})$ for $M_{\rm
PBH}=10^{-6}M_{\odot}$ $(10^{2}M_{\odot})$, the first baryonic structure formation
will be different. Fig. \ref{figMhalo_B} shows the minimal dark matter mass for the
critical case $T_{\rm m}=T_{\rm vir}$ with different $\eta f_{\rm
PBH}$ and the characteristic radiation $E_{X}=10$ keV. Different
from the dark matter annihilation heating case (Fig.
\ref{figMhalo_A}), the minimal dark halo mass dramatically increase
after $z_{m}$. For $z_{\rm vir}=20$, the minimal dark haloes increase
to $>10^{6}M_{\odot}$ for $\eta f_{\rm PBH}\geq10^{-8}$. Moreover,
if we combine the annihilation before $z_{m}$ with the X-ray
emission after $z_{m}$, the minimal dark halo mass for $T_{\rm
m}=T_{\rm vir}$ can be even larger.

Moreover, we expect the gas accretion onto PBHs in the dark halo environment above the critical mass in Fig.
\ref{figMhalo_B} will be slightly different from that in the ambient IGM,
because the baryon gas is denser within a dark halo than the
ambient IGM, which leads to a different accretion rate and baryon
fraction inside UCMHs compared to the IGM-located-UCMHs. The
accretion rate $\dot{M}_{\rm UCMH}$ and baryon fraction $f_{b}$ inside a
PBH host UCMH should be higher than those outside the halo.
Therefore it is more difficult for X-rays from PBHs to pass through the host UCMH without
absorption. The critical redshift $z_{m}^{\rm halo}$ for X-rays
escaping from the UCMH baryonic environment should be slightly
delayed inside the halo than that in the background $z_{m}^{\rm
halo}<z_{m}^{\rm bkgd}$. It is possible that in a period of time
that X-ray energy injection and deposition within a halo is mainly
from the background even after virialization.

\begin{figure}
\centerline{\includegraphics[width=100mm]{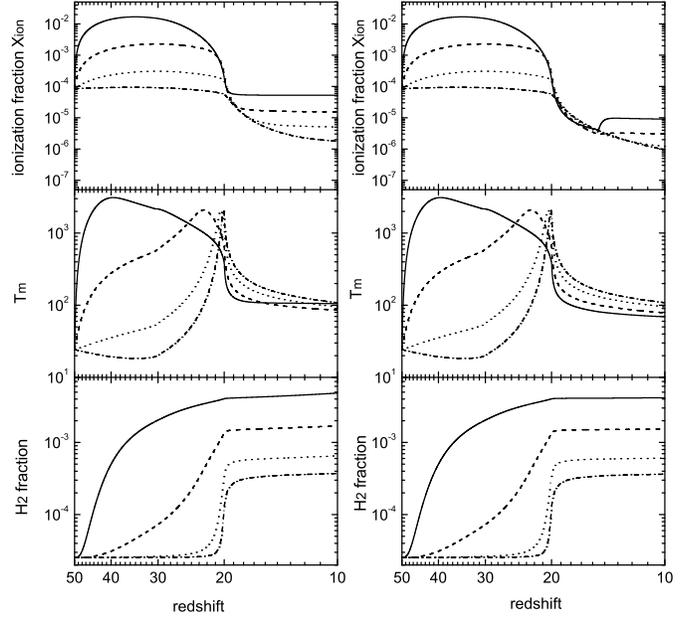}}
\caption{Effects of gas accretion onto PBH inside a $10^{7}M_{\odot}$ isothermal
halo with $z_{\rm vir}=20$, $z_{m}^{\rm bgd}=50$, $z_{m}^{\rm halo}=30$ (left panel) and
$z_{m}^{\rm UCMH}=15$ (right panel).
The characteristic $E_{X}=10$ keV with $\eta f_{\rm PBH}=10^{-7}$ (solid lines),
$\eta f_{\rm PBH}=10^{-8}$ (dashed line, $\eta f_{\rm PBH}=10^{-9}$ (dotted lines), $\eta f_{\rm PBH}=10^{-10}$ (dash-dotted lines).
We choose a more massive halo $10^{7}M_{\odot}$
because more massive haloes are favors for the structure formation in this case.
In the left panel we take $\frac{dn_{\rm UCMH}}{dM_{\rm halo}}$ uniformly and right panel
$\frac{dn_{\rm UCMH}}{dV_{\rm halo}}$ uniformly.}\label{figSPBH}
\end{figure}

The evolution of the baryonic structure inside a $10^{7}M_{\odot}$
isothermal dark halo, as an example, is showed in Fig.
\ref{figSPBH}. We consider two models: the number density of PBH
host UCMHs being uniformly distributed per halo mass as mentioned in
Section \ref{secStruc01}; also the uniformly distributed UCMHs per
halo volume inside the halo as $\frac{dn_{\rm UCMH}}{dV_{\rm
halo}}\propto const$. Different UCMH distribution inside the halo
would give different baryonic evolution. The local energy deposition
for the UCMHs' uniform distribution per halo volume is written as
\begin{eqnarray}
\epsilon_{\rm loc,acc}(z)\simeq\frac{L_{\rm
acc}\sigma_{\rm tot}E_{X}\rho_{\rm core}(1-f_{\chi})}{\mu m_{p}f_{\chi}V_{\rm
halo}}\left[2R_{\rm core}-\frac{R_{\rm core}^{2}}{R_{\rm tr}}\right],\label{structure02}
\end{eqnarray}
where $L_{\rm acc}$ is the total X-ray luminosity due to gas
accretion onto PBHs inside the first baryonic object. The main
results in Fig. \ref{figSPBH} is very similar to those of dark
matter annihilation in Fig. \ref{figShalo}. For low virialization
redshift $z_{\rm vir}=20$, the gas temperature is cooler
for higher X-ray luminosity, but $T_{\rm m}$ inside the halo only
changes by a factor of 2 to 4 for a four orders of magnitude change
in the X-ray luminosity. An more obvious change of chemical
quantities $(x_{\rm ion},f_{\rm{H}_{2}})$ than the temperature
change occurs for different X-ray luminosity within the halo. This
means the effects of X-ray emission form PBHs on the gas evolution inside a halo is very small. PBHs
which uniformly distribute per halo volume gives a
slighter cooler gas within the halo, as well as lower $x_{\rm ion}$
and $f_{\rm H_{2}}$ than the uniformly distributed UCMHs per halo
mass. That is because the latter model makes the average distance of
UCMHs to be closer to the halo center, and gives more effects to
change the gas properties at the center.

In summary, UCMH radiation including both annihilation and PBH gas
accretion enhances the baryon chemical quantities such as $x_{\rm
ion}$ and $f_{\rm H_{2}}$ inside dark matter haloes which above the
minimal halo mass for $T_{\rm m}=T_{\rm vir}$, but the impact of UCMH
radiation on the temperature of first baryonic objects is small (by
a factor of several), which shows the change of first baryonic
structure formation due to UCMH radiation is less important than the
$\textrm{H}_{2}$ cooling and dark halo virialization time.
However, the new chemical conditions provided by UCMH radiation can
be more important to affect the later gas collapse and first star
formation after the first baryonic object formation, because more
abundant $\textrm{H}_{2}$ and electrons acting as the cooling agents
can cool the gas more efficient during the gas collapse process, and
provide a lower fragmentation mass scale and first star mass (e.g., \citealt{Stacy07}). As we
mainly focus on the first baryonic structure formation and evolution, the detailed calculation
of first star formation due to the changed gas chemical components
should be investigated more detailed in the future.

\section{Discussion}\label{SecDis}

\subsection{Status of UCMH Radiation in Reionization, Other Sources}\label{SecDis01}
A variety of cosmological sources can reionize and heat the IGM at
different redshifts before $z\simeq6$. So far we showed that UCMHs,
even merely occupy a tiny fraction of total dark matter mass, provide a new
gamma-ray background for gas heating and ionization. Also,
the X-ray emission from the accreting PBHs could change the
IGM gas evolution history dramatically after $z_{m}\ll1000$,
where the value of $z_{m}$ depends on the masses of PBH, host
UCMHs and dark matter particles. Furthermore, we investigate that
both dark matter annihilation and X-ray emission from UCMHs can
dominate over the annihilation of extended dark matter halo. Therefore
UCMHs are also an important energy source in dark matter haloes before the
first star formation. In this section we briefly review all candidate energy
sources during the Universe reionization era $10\leq z<z_{\rm eq}$.
In particular, we emphasize the importance of UCMH
radiation among all of these sources in different times.

In this paper we focus on the heating and ionization processes after the last
scattering epoch $z\sim1000$, but some interesting effects can be
produced by primordial energy sources at earlier time $z>1000$. For example,
cosmic gas heating and CMB spectral distortion at $z_{\rm rec}<z<z_{\rm eq}$
produced by PBH gas accretion can be used to constrain the PBH
abundance that $f_{\rm PBH}\leq10^{-8}$ for $M_{\rm PBH}\geq10^{3}M_{\odot}$ in the absence
of UCMH annihilation (\citealt{Ricotti08}). However, as showed in Section \ref{secAccPBH},
UCMH annihilation can be more important than PBH gas accretion in the very earlier
Universe even though UCMHs are just beginning to grow at that time. Hotter
cosmic gas heated by dark matter annihilation suppresses the PBH gas accretion to
becomes the dominated sources to distort CMB, and even changes the cosmic
recombination process as showed
in Fig. \ref{figIGMhalo}. Compton y-parameter is likely to be used to constrain
the UCMH abundance based on the annihilation scenario in future work.

The influence of dark matter annihilation or decay at $z\leq1000$ on
the IGM during the reionization era has been discussed by many authors
(\citealt{Chen04}; \citealt{Hansen04}; \citealt{Pierpaoli04};
\citealt{Mapelli05}; \citealt{Padmanabhan05}; \citealt{Zhang06};
\citealt{Mapelli06}; \citealt{Ripamonti07}; \citealt{Yuan10};
\citealt{Chluba10}). Some authors also use the observation
data to constraint the cross section of the dark matter interaction
(\citealt{CIP09}; \citealt{Galli09}; \citealt{Slatyer09};
\citealt{Kanzaki10}). The basic assumption is that the dark matter
distribution is smooth and homogenous at $z\geq100$. However,
the annihilation power can be strongly increased by UCMHs in the early Universe,
as what we have discussed in this paper.

The next commonly suggested sources of ionization and heating is the
first dark objects (dark haloes), which formed approximately at
$z\leq100$. Dark matter haloes enhanced the overall cosmic dark matter
annihilation density due to the dark matter concentration
in haloes (\citealt{Iliev05}; \citealt{Oda05};
\citealt{Ciardi06}; \citealt{Chuzhoy08}; \citealt{Myers08};
\citealt{NS08}; \citealt{BH09}; \citealt{NS09}; \citealt{NS10}). The
mass distribution of dark haloes varies from very low mass at
($\sim10^{-6}M_{\odot}$) to high mass ($\sim10^{12}M_{\odot}$)
(\citealt{Green05}; \citealt{Diemand05}; \citealt{Hooper07}),
depending on the different damping scales due to different dark
matter models (\citealt{Abazajian01}; \citealt{Boehm05}), as well as
the mass-halo function (\citealt{Press74}; \citealt{ST99}). If UCMHs
collapse with the homogenous dark matter together, the UCMH
annihilation flux could still dominate over the total annihilation
flux within the dark matter haloes, at least in massive dark haloes
with $f_{\rm UCMH}M_{\rm DM}\gg m_{h}$. However, remember that
small dark matter haloes contribute to a significant part
of the total annihilation rate after structure formation. The
profiles of the earth-mass dark matter haloes and the gamma-ray flux due
to annihilation have also been studied recently
(\citealt{Diemand05}; \citealt{Ishiyama10}). Similar to the UCMH
emission, a large enhancement of annihilation signal is also
expected due to the emission from the dark matter subhaloes as the
remnant of structure formation at $z<60$. Whether UCMHs or small haloes are
more important for ionization and heat after $z\sim60$ should be
investigated in the future.

The following ionization sources are the accreting PBHs, which locate
in their host UCMHs, as mentioned in Section \ref{secAccPBH}. PBHs with
host UCMHs lead to a faster accretion than naked PBHs, but
also absorb the X-ray emission due to baryon accumulation within the
UCMH. The PBH accreting could only be more important than UCMH annihilation
at $z\leq z_{m}\ll1000$ with sufficient abundance and radiation efficiency.
Keep in mind that the X-ray emission here is
from PBHs, or say, the PBH-UCMH systems, which paly an earlier
role than the so-called accreting ``first black holes (BHs)'', which
are the remnants of first stars at $z\sim15$.

As mentioned in Section \ref{secStruc}, dark matter annihilation or
X-ray emission affects the baryonic structure formation and
evolution. Also, they affect the process of first
star formation. The standard first star formation carried out at
$z\sim20$ (\citealt{Abel02}; \citealt{Bromm09}), but the first star
forming history can be affected by the primordial magnetic fields
(\citealt{Tashiro06}), or by extended dark matter annihilation in
the halo (so-called ``dark star'', see \citealt{Spolyar08};
\citealt{Spolyar09}). Previously it was said that the first stars
gave the first light to end the cosmic ``dark age'', that cannot be
true if exotic sources such as dark matter annihilation and accreting PBHs are
included.

Next ionization sources are more familiar to us. First stars emitted
UV light and produced the ``ionized bubbles'', which could directly
partially ionize the Universe at $z<20$, or affected the coming
formation of next generation stars and later galaxy formation (e.g.,
\citealt{Haiman97}; \citealt{Wyithe03}; \citealt{Shull08};
\citealt{Whalen10}). The death of first stars,  produced the ``first
generation BHs'', which emitted X-rays and ionized the Universe at
$z\sim15$ or even closer (\citealt{Cen03}; \citealt{Ricotti04};
\citealt{Madau04}; \citealt{Ripamonti07}; \citealt{Thomas08}). The
reionization process was completed after galaxy formation, as
galaxies are generally considered the main candidates for the
reionization of the Universe at $z\sim6$ (\citealt{Meiksin09}, and
references therein).

Future work that can be done includes studying the heating and ionization
processes at $z>1000$ due to annihilation, comparing the total
annihilation rate from small dark haloes ($10^{-6}M_{\odot}<M_{\rm
DM}<f_{\rm UCMH}^{-1}m_{h}$ as we mentioned above) with that from
UCMHs, and distinguish the impacts of different ionization sources
using the CMB polarization anisotropies and 21 cm spectra observational constraints.
Actually, CMB polarization and hydrogen 21 cm line are
powerful potential probes of the era of reionization to constrain
the early energy sources. The high multipoles of polarization anisotropies may
be able to distinguish UCMHs from small dark structures formed at $z<100$, and
further constrain the UCMH and PBH abundances.

\subsection{Different UCMH Profiles}\label{SecDis02}

\begin{figure}
\centerline{\includegraphics[width=100mm]{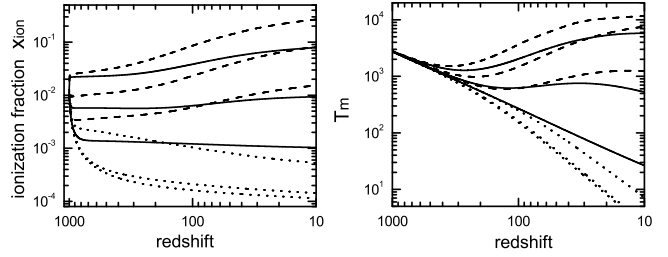}}
\caption{An example of IGM evolution caused by UCMH annihilation with different UCMH
profiles. In order to see the effects more clearly, UCMH fraction is taken as $f_{\rm UCMH}=10^{-4}$.
The same group of lines from the top down show the results for
$m_{\chi}c^{2}=1$ GeV, 10 GeV and 100 GeV with $\rho\propto r^{-9/4}$ (solid lines), $\rho\sim r^{-3}$ (dashed lines)
and $\rho\sim r^{-1.5}$ (dotted lines).}\label{figProfile}
\end{figure}

Remember that in Section \ref{secanni} although several UCMH
annihilation rate due to various profiles were given in Fig.
\ref{figratio}, we choose the UCMH profile as $\rho\propto r^{-9/4}$
with a cut off at $\rho_{\rm max}\propto(t-t_{\rm i})^{-1}$.
Such a profile is based on the analytical solution of the radial infall onto a
central overdensity (\citealt{Bert85}). A shallower density profile $\rho(r)\propto
r^{-1.5}$ which is given if the central accretor is a black
hole, or a steeper profile $\rho(r)\propto r^{-3}$ simulated by
\cite{Mack07}, will change the total annihilation luminosity of a
UCMH significantly. Fig. \ref{figProfile} gives an example of the
different annihilation luminosity due to different density profiles
in an entire UCMH. More concentrated dark matter distribution in a steeper profile
leads to much higher total annihilation rate within the UCMH,
because the center region of a UCMH contributes to most part of the total
annihilation rate. However, we conclude that the overall cosmic annihilation luminosity density equation
(\ref{halo08}) will not be changed too much for two reasons. First of all,
the $\rho\propto r^{-3}$ profile usually appears in the outer region of
a UCMH, but the change of the UCMH density profile at the
outer region $r\gg r_{\rm cut}$ will not dramatically change the total annihilation rate,
the density distribution at the central region is crucial to determine the
total annihilation rate. Second, the contribution of PBH host UCMHs (with the profile
$\rho(r)\propto r^{-1.5}$ near the center) to the
overall cosmic annihilation should be much
less important than the initial overdensity seeded UCMHs ($\rho(r)\propto r^{-2.25}$),
both due to their shallower inner profile and
the much lower abundance $f_{\rm PBH}\ll f_{\rm UCMH}$.
The annihilation luminosity should still be taken into account if the dark matter particle inner
trajectory is high eccentric with a much closer pericenter than the
cut off radius as in equation (\ref{anni04}), but it is still lower than
the luminosity of the overdensity seeded UCMHs as showed in Section \ref{secanni} (\citealt{Lacki10}).

\subsection{Extended Radiation Spectral Energy Distribution}\label{SecDis03}

In Section \ref{secIonHeat} the products of two dark
matter particles annihilation are assumed to be two gamma-ray photons both with
energy $m_{\chi}c^{2}$, and X-ray photons emitted from PBH host
UCMHs are characterized with the single energy reflecting
the center PBH mass $E_{X}\simeq3$ keV $(M_{\rm
PBH}/M_{\odot})^{-1/4}$. The simplified
treatment with a monochromatic (i.e., $\delta$-function) spectrum of local UCMH emissivity
is a good approach to demonstrate the crucial UCMH effects on the IGM evolution depending on
the most crucial parameters such as UCMH and PBH abundances $f_{\rm
UCMH}$ and $f_{\rm PBH}$, as well as dark matter particle mass
$m_{\chi}$. In this section we will experiment other extended photon
spectral energy distributions (SEDs) for UCMH radiation rather than
$\delta$-function, and study the dependence of UCMH emissivity with
SED. We will see that more elaborate considerations can
quantitatively change the ionization results, but will not change
the basic conclusions of UCMH radiation qualitatively. We first adopt a power law
spectrum as $F(E,z)\propto E^{-a}$, and then discuss another
spectrum as $F(E,z)\propto E^{b}\exp(-dE)$, which are two most
commonly used SEDs for annihilation and BH X-ray emission.

First of all we give an analytic calculation for UCMH radiation with
locally monochromatic spectrum, based on an approximation that
the optical depth described as equation (\ref{baseq04}) can be
neglected $\tau\ll1$. This condition is applied to the spectrum
$E_{\gamma}(E_{X})\gg10$ keV. Under the approximation $\tau\ll1$ equation
(\ref{baseq03}) in Section \ref{secIonHeat} can be simplified as
\begin{equation}
\frac{dn}{dE_{\gamma}}\simeq\frac{1}{E_{0}E_{\gamma}}\frac{c}{H_{0}\sqrt{\Omega_{m}}}
\left(\frac{E_{\gamma}}{E_{0}}\right)^{\frac{11}{2}-\zeta}\frac{l(z)}{(1+z)^{3/2}}.\label{dis021}
\end{equation}
where the emitted photon number $n$ and luminosity $l(z)$ can be applied to both
annihilation $(n_{\rm ann}, l_{\rm ann}(z))$ or X-ray luminosity
$(n_{\rm X}, l_{\rm acc}(z))$, $E_{0}$ for $E_{\chi}$ or $E_{X}$
respectively, and $l(z)\propto(1+z)^{\zeta}$ with
$\zeta\approx4$ for annihilation and $\zeta\approx3$ for X-ray
emission. We take the interaction cross section between photons and
the IGM gas as $\sigma_{\rm tot}(E)\propto E^{-k}$. The total
energy deposition per second per volume equation (\ref{baseq05}) is
integrated as
\begin{equation}
\epsilon_{\delta}(z)=\frac{c}{H_{0}\sqrt{\Omega_{m}}}\left(\frac{13}{2}-\zeta-k\right)^{-1}l(z)n_{A}(1+z)^{3/2}\sigma_{\rm tot}(E_{0}),\label{dis022}
\end{equation}
where the subscript $\delta$ is marked for monochromatic SED as
in equation (\ref{baseq05}). For high energy gamma-ray photons we
can take $k\approx2$ for Klein-Nishina cross section but for X-ray
photons we take $k\approx0$ for $E_{X}\gg1$ keV. Another thing we
mention is that ionization rate $I(z)\propto\epsilon(z)$ for $x_{\rm
ion}\ll1$, so we only track $\epsilon(z)$ based on different SEDs.

The first extended SED is $F(E,z)\propto E^{-1}$ for $E_{1}\leq
E\leq E_{2}$, which gives a spectral number density for $\tau\simeq0$
\begin{equation}
\frac{dn}{dE_{\gamma}}\simeq\frac{c}{H_{0}\sqrt{\Omega_{m}}}\left(\frac{2}{13-2\zeta}\right)\frac{l(z)}{\ln\Lambda}(1+z)^{-3/2}E_{\gamma}^{-2}\label{dis023},
\end{equation}
where $\Lambda=E_{2}/E_{1}$. Note that the number spectrum equation (\ref{dis023}) is
different from the equation (\ref{dis021}) in the monochromatic SED case. If we consider dark
matter annihilation $k\simeq2$, the total energy deposition for UCMH
annihilation is written as
\begin{eqnarray}
\epsilon(z)&\simeq&\frac{c}{H_{0}\sqrt{\Omega_{m}}}\left(\frac{2}{13-2\zeta}\right)\frac{l_{\rm ann}(z)}{k\ln\Lambda}n_{A}(1+z)^{3/2}\nonumber\\
&&\times\left(\frac{E_{0}}{E_{1}}\right)^{k}
\sigma_{\rm tot}(E_{0})\left(\frac{1+z_{\rm eq}}{1+z}\right)^{k}\label{dis024}.
\end{eqnarray}
Compared with equation (\ref{dis022}), the most obvious difference
is the additional two factors $(E_{0}/E_{1})^{k}$ and $(1+z_{\rm
eq})^{k}/(1+z)^{k}$ significantly increase the annihilation
energy density if $E_{0}\gg E_{1}$. However
This energy amplification for power law SED might be overestimated
as we take the lower limit of the energy (\ref{baseq05}) as $E_{\rm
eq}=E_{1}(1+z)/(1+z_{\rm eq})$, i.e, the Universe at redshift $z$
can receive the emission from the matter-radiation equality era. A
more general expression $(E_{1}/E_{i})^{k}$ can be used instead of
$(1+z_{\rm eq})^{k}/(1+z)^{k}$ with $E_{i}$ being the threshold
energy photons from UCMH radiation, then the amplified factor
compared to equation (\ref{dis022}) can be written as
$(E_{0}/E_{i})^{k}\gg1$, no direct relation with $E_{1}$ and
$E_{2}$. On the other hand, for the X-ray emission from a PBH host
UCMH, as $k\approx0$ and the cross section is more or less the
Thomson cross section, then the energy density is written as
\begin{eqnarray}
\epsilon(z)&\simeq&\frac{c}{H_{0}\sqrt{\Omega_{m}}}\left(\frac{2}{13-2\zeta}\right)l_{\rm acc}(z)n_{A}(1+z)^{3/2}\sigma_{T}\nonumber\\
&&\times\left[1+\ln\left(\frac{E_{1}}{E_{i}}\right)/\ln\Lambda\right]\label{dis025}
\end{eqnarray}
In this case the energy density can be increased compared to the
monochromatic spectrum if $E_{1}>E_{i}$, or decreased for
$E_{1}<E_{i}$. But the logarithmic enhancement $\ln(E_{1}/E_{i})$ in
X-ray emission case is less significant than the annihilation case.

For a more general power law spectrum $F(E,z)\propto E^{-a}$
($E_{1}\leq E\leq E_{2}$) with $a>1$, the number density spectrum
and energy deposition are calculated as
\begin{equation}
\frac{dn}{dE_{\gamma}}\simeq\frac{c}{H_{0}\sqrt{\Omega_{m}}}\left(\frac{a-1}{a+\frac{11}{2}-\zeta}\right)(1+z)^{-3/2}
\left(\frac{E_{1}}{E_{\gamma}}\right)^{a}\frac{l(z)}{E_{1}E_{\gamma}},\label{SEDpower01}
\end{equation}
and
\begin{eqnarray}
\epsilon(z)&\simeq&\frac{c}{H_{0}\sqrt{\Omega_{m}}}\frac{l(z)}{a+k-1}\left(\frac{a-1}{a+\frac{11}{2}-\zeta}\right)n_{A}(1+z)^{3/2}\nonumber\\
&&\times\sigma_{\rm tot}(E_{0})\left(\frac{E_{0}}{E_{1}}\right)^{k}\left(\frac{E_{1}}{E_{i}}\right)^{a+k-1}\nonumber\\
&\propto&\epsilon_{\delta}(z)\times\left(\frac{E_{0}}{E_{1}}\right)^{k}\left(\frac{E_{1}}{E_{i}}\right)^{a+k-1}.\label{SEDpower02}
\end{eqnarray}
If we do not focus on the linear changed factors such as
$(a-1)/(a+\frac{11}{2}-\zeta)$, mostly the power low spectrum will significantly increase the energy deposition
$\epsilon(z)$ as well as the ionization rate
$I(z)$ (and even a dramatic increase in some
cases) by a factor of $(E_{0}/E_{k})^{k}(E_{1}/E_{i})^{a+k-1}$.

A black-body-like or say multicomponent spectral distribution
$F(E,z)\propto E^{b}\exp(-dE)$ with a peak $E_{0}$ and $c>0$ can be
approximately written as $F(E,z)\propto x^{b}$ for $x<1$ and
$F(E,z)\propto \exp(-dx)$ for $x>1$, where $x=E/E_{0}$. In this case we
find the number density spectrum as
\begin{eqnarray}
\frac{dn}{dE_{\gamma}}&\simeq&\frac{cA}{H_{0}\sqrt{\Omega_{m}}}\frac{l(z)}{E_{0}E_{\gamma}}\frac{(1+z)^{-3/2}}{\frac{11}{2}-b-\zeta}\nonumber\\
&&\times\left[\left(\frac{E_{\gamma}}{E_{0}}\right)^{b}+(f(d)-1)\left(\frac{E_{\gamma}}{E_{0}}\right)^{\frac{11}{2}-\zeta}\right],
\label{SEDBB}
\end{eqnarray}
where $A$ is an algebraic factor $A=[(1+b)^{-1}+d^{-1}]^{-1}$, and
$0<f(d)<1$ can be obtained numerically, which is not important for
the following discussion. The terms in
number density spectrum are proportional to $E_{\gamma}^{b}$ or
$E_{\gamma}^{11/2-\zeta}$. According to equation (\ref{SEDBB}), the
integrated energy density $\epsilon(z)$ is expected to be similar as
equation (\ref{dis022}) for $b>k-1$. Otherwise the enhancement
should be
$\epsilon(z)\propto\epsilon_{\delta}(z)(E_{0}/E_{1})^{k-b-1}$. A
steep $x^{c}$ with $c>1$ for annihilation $k\sim2$ and all the
black-body-like SED for X-rays $k\simeq0$, are more or less similar
to the $\delta$-function SED at $E_{0}$.

Now we study the case of $E_{X}\ll10$ keV for X-ray emission from very massive PBHs.
Photon absorption in the IGM is important for $E_{X}\sim1$
keV. The mean free path of X-ray photons describing by redshift change $\Delta
z$ as
\begin{equation}
\tau\sim\Delta z\frac{c}{H_{0}\sqrt{\Omega_{m}}}(1+z)^{-5/2}n_{A}(1+z)^{3}\sigma_{\rm tot}(\textrm{1\,keV})\left(\frac{E_{X}}{1\,\textrm{keV}}\right)^{-k},
\label{SED04}
\end{equation}
or we have
\begin{equation}
\Delta z\sim0.40(1+z)^{-1/2}\left(\frac{E_{X}}{1\,\textrm{keV}}\right)^{k},\label{SED05}
\end{equation}
where $k\simeq3.3$ for the photonionization cross section. Similarly
as the former calculation, they energy density for a
$\delta$-function spectrum is
\begin{equation}
\epsilon_{\delta}(z)\sim\frac{0.40}{H_{0}\sqrt{\Omega_{m}}}l_{\rm acc}(z)n_{A}\left(\frac{E_{0}}{1\,\textrm{keV}}\right)^{k}\sigma_{\rm tot}(E_{0}).
\label{SED06}
\end{equation}
Note that now we have a shallower density evolution
$\epsilon_{\delta}(z)\propto (1+z)^{3}$ compared with the
transparently propagation for the case of equation (\ref{dis024}) as
$\epsilon_{\delta}(z)\propto (1+z)^{\zeta+3/2}$. For a power law SED
$F(E,z)\propto E^{-a}$ ($E_{1}\leq E\leq E_{2}$), we obtain the
energy density as
\begin{eqnarray}
\epsilon(z)&\sim&\frac{0.40c}{H_{0}\sqrt{\Omega_{m}}}l_{\rm acc}(z)n_{A}\sigma_{\rm tot}(E_{0})
\left(\frac{E_{0}}{1\,\textrm{keV}}\right)^{k}\left(\frac{E_{1}}{E_{0}}\right)^{a-1}\nonumber\\
&&\propto\epsilon_{\delta}(z)\left(\frac{E_{1}}{E_{i}}\right)^{a-1}.\label{SED07}
\end{eqnarray}
Therefore the energy density is enhanced for $E_{1}>E_{i}$, but the
enhancement is less significant compared with that of equation
(\ref{SEDpower02}) for the same spectrum index $a$.

For a brief summary, the strength of UCMH radiation depends on its
extended spectral energy distribution, which will increase the
properties of IGM ionization $I(z)$ and heating $\epsilon(z)$.
Compared to the basic results with the monochromatic spectrum
$E_{\chi}$ for annihilation or $E_{X}$ for X-ray emission, power law
spectrum $\propto E^{-a}$ with $a>0$ can increase the energy density
effectively, but black-body-like spectrum is more like the
monochromatic SED case. For locally heating $E_{X}\sim1$ keV,
the heating increases less significant than the transparently
propagation case $\tau\ll1$. Also, power law spectrum changes
$\epsilon(z)$ and $I(z)$ more significant for annihilation than
X-ray radiation.

\subsection{More Massive UCMHs Inside First Dark Haloes}\label{SecDis04}

In Section \ref{secStruc} we assume that the number of UCMHs inside
a dark halo is so huge that UCMHs in the halo are uniform
distributed per halo mass $\frac{dn_{\rm UCMH}}{dM_{\rm
DM}}=const.$ or per volume $\frac{dn_{\rm UCMH}}{dV_{\rm
DM}}=const.$  This assumption
can be invalid if the mass seed of a single UCMH $m_{h}$ is
comparable to $f_{\rm UCMH}M_{\rm DM}$. In this case the position of each UCMH
is important to determine the energy deposition within the halo. Usually X-ray emission can be neglected
in a massive UCMH due to photon trapping (Section \ref{secAccPBH3}),
we only focus on the UCMH annihilation. Equation (\ref{deposition2}) in
Section \ref{secStruc02}is invalid for massive UCMHs $m_{h}\sim f_{\rm UCMH}M_{\rm DM}$.
In this case $\epsilon_{\rm loc,iso}$ in equation (\ref{deposition2}) should be
calculated as the summation of individual UCMHs
\begin{equation}
\epsilon_{\rm loc,UCMH}(z)=\frac{\sigma_{\rm tot}(1-f_{\chi})}{\mu m_{p}f_{\chi}}\sum_{i}\frac{L_{\rm{UCMH},\it{i}}\rho(r_{i})}{4\pi r_{i}^{2}}\label{deposition3},
\end{equation}
where $L_{\rm{UCMH},\it{i}}$ and $r_{i}$ are the annihilation
luminosity and position of the $i$-th UCMH within the halo.

\begin{figure}
\centering\includegraphics[width=80mm]{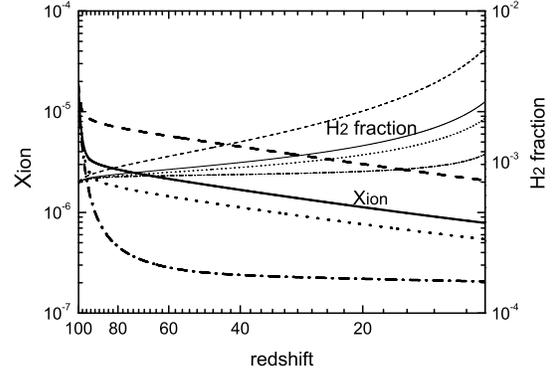}
\caption{Ionization fraction $x_{\rm ion}$ (thick lines) and molecular hydrogen fraction
H$_{2}$ (thin lines) in a $10^{6}M_{\odot}$ isothermal halo with a single massive UCMH
located at $R_{\rm core}$ (dashed lines), $2R_{\rm core}$ (dotted lines) and $10R_{\rm core}$ (dash-dotted lines)
from the center of the halo. The solid lines shows the results given by many small UCMHs which number density is proportional to
the halo density. Both single UCMH or small UCMHs is adopted a total UCMH fraction as $f_{\rm UCMH}=10^{-4}$, $z_{\rm vir}=100$
and $m_{\chi}c^{2}=10$ GeV.}\label{figsingle}
\end{figure}

An extreme case is that there is only one massive UCMH inside a dark halo, the energy
deposited in the gas at the center of the halo can be obtained by
equation (\ref{deposition3}) in Section \ref{secStruc01}. Remember
that for a uniformly distributed UCMHs with a same total mass of the
single UCMH inside a halo we use equation (\ref{deposition2}). The
ratio factor between (\ref{deposition3}) and (\ref{deposition2}) is
\begin{eqnarray}
\textrm{factor}&=&\left(\frac{R_{\rm core}}{r_{0}}\right)^{4}\left(\frac{3R_{\rm tr}}{R_{\rm core}}-2\right)
\left[4-\left(\frac{R_{\rm core}}{\rm tr}\right)^{3}\right]^{-1}\nonumber\\
&&\simeq\frac{3}{4\xi}\left(\frac{R_{\rm core}}{r_{0}}\right)^{4},\label{ratio01}
\end{eqnarray}
where $r_{0}$ is the location of the single massive UCMH from the
halo center. If we take $R_{\rm core}=\xi R_{\rm tr}$ with
$\xi\ll1$, the factor can be approximately written as
$(3/4\xi)(R_{\rm core}/r_{0})^{4}$. For a single UCMH located close
to the isothermal case $r_{0}\simeq R_{\rm core}$, the factor
(\ref{ratio01}) is $3/(4\xi)$. For a single UCMH located at the
turnaround radius $r_{0}\simeq R_{\rm tr}$ equation (\ref{ratio01})
becomes $\simeq3\xi^{3}/4$. Taking a typical value $\xi\sim0.1$, the
energy deposition $\epsilon_{\rm loc}$ contributed by a single UCMH
varies from $\sim10$ higher to $\sim 10^{-3}$ lower compared to that
contributed by the same total mass but uniformly distributed small
UCMHs. Fig. \ref{figsingle} gives the gas $x_{\rm ion}$ and $f_{\rm
H_{2}}$ inside a $10^{6}M_{\odot}$ isothermal halo with a single
UCMH located in difference locations. UCMH annihilation on the halo
center will be less important for further located UCMH $>2R_{\rm
core}$. The uniformly distributed UCMHs are more or less equivalent
to a singe UCMH located at approximately $\sim1.5R_{\rm core}$. As
an example in this figure, we take the UCMH fraction as $f_{\rm
UCMH}=10^{-4}$ and $m_{\chi}c^{2}=10$ GeV. More massive dark matter
particles or less UCMH fraction will give a faster decrease in
energy deposition with the single UCMH moving outward the halo. We
also check the importance of a single UCMH's position on the gas
temperature inside a halo, but we see that the effect can be
neglected for the temperature.

On the other hand, the energy deposition by a single UCMH compared
with the volume uniformly distributed small UCMHs with a same total
mass can be amplified by a factor
\begin{eqnarray}
\textrm{factor}&\simeq&\frac{1}{6}\left(\frac{R_{\rm core}}{r_{0}}\right)^{4}\left(\frac{R_{\rm tr}}{R_{\rm core}}\right)^{3}
\left(1-\frac{R_{\rm core}}{R_{\rm tr}}\right)\nonumber\\
&\simeq&\frac{1}{6\xi^{3}}\left(\frac{R_{\rm core}}{r_{0}}\right)^{4},\label{ratio02}
\end{eqnarray}
which gives an enhancement factor about
$\sim1/(6\xi^{3})\sim1/\xi^{2}$ for $r_{0}=R_{\rm core}$ and a
weaker factor $\xi/6$ for $r_{0}=R_{\rm tr}$. These results show
that the massive single UCMH is always more important for energy
deposition compared with the volume uniformly distributed UCMHs.

Therefore we conclude that different UCMH distribution with a same
total mass but different individual mass can change the energy
deposition and structure evolution inside a halo. More concentrated
distribution towards the halo center or a closer located single halo
near the center gives a more significant effect on the gas ionization and heating
at the halo center.

\section{Conclusions}\label{SecCon}

Ultracompact Minihaloes (UCMHs) have been proposed as the primordial
dark matter structures which formed by dark matter accreting onto
initial overdensity or primordial black holes (PBHs) after
matter-radiation equality $z_{\rm eq}\simeq3100$ (\citealt{Mack07};
\citealt{Ricotti09}). The key difference between UCMHs and the
first dark halo structures is that, UCMHs are seeded by primordial
density perturbations produced in the very early Universe such as
the phase transition epoches ($10^{-3}\leq\delta\leq0.3$ for the initial
overdensity or $\delta>0.3$ for PBHs), so they can grow shortly
after $z\sim z_{\rm eq}$. The radiation from UCMHs in the early
Universe includes dark matter annihilation from all UCMHs, and X-ray
emission from gas accretion onto PBHs. In this paper we investigate
the influence of UCMH radiation on the early ionization and thermal
history of the intergalactic medium, and the following evolution of
the first massive baryonic objects. Our conclusions are as follows.

1. UCMH annihilation can totally dominate over the homogenous dark
matter background annihilation, and provide a new gamma-ray
background even for a tiny UCMH fraction $f_{\rm
UCMH}=\Omega(z_{\rm eq})/\Omega_{\rm DM}\sim2.2\times10^{-15}m_{\chi,100}^{-2/3}(1+z)^{2}$ with
$m_{\chi,100}=m_{\chi}c^{2}/100$ GeV. We
conclude that the influence of dark matter annihilation on
the IGM evolution can be significantly enhanced when we include UCMHs
besides the homogenous dark matter background. In most cases
the UCMH annihilation had been the dominated sources of ionization and heating gas since matter-radiation equality epoch,
until the X-ray emission from PBHs or large scale structure
formation become important at $z\leq100$.

2. The impact of UCMH annihilation on the IGM can be approximately estimated by the
quantity $m_{\chi,100}^{-1}f_{\rm UCMH}$. The threshold of UCMH abundance $f_{\rm UCMH}$ to affect the IGM
evolution by dark matter annihilation is
$m_{\chi,100}^{-1}f_{\rm UCMH}>10^{-6}$. After matter-radiation equality
epoch, the IGM ionization fraction $x_{\rm ion}$ can be increased
from $x_{\rm ion}\sim10^{-4}$ in the absence of any energy
injections to an upper bound $x_{\rm ion}\sim0.1$, and the IGM temperature from
the adiabatical cooling $T_{\rm m}\propto(1+z)^{2}$ to a maximum value
$T_{\rm m}\sim 5000$ K for the upper bound case $m_{\chi,100}^{-1}f_{\rm UCMH}\sim10^{-2}$,
which is constrained using the CMB optical depth at late times $z<30$.
UCMH annihilation is able to significantly increase the Thomson optical depth $\tau\geq0.1$
in the early Universe $z\gg30$, which is unrelated with the measured CMB optical depth at $z<30$. The UCMH
annihilation luminosity is based on the UCMH profiles, where we take
$\rho\propto r^{-2.25}$ from literature (\citealt{Bert85}), steeper (shallower)
profiles decrease (increase) the allowed upper limit of $f_{\rm
UCMH}$, but the variations of the overall IGM chemical and thermal quantities
should not be changed too much, because the fraction of UCMHs with a
profile $\rho\propto r^{-1.5}$ as the PBH hosts are very small, and
$\rho\propto r^{-3}$ occurs only in the halo outskirts $r\gg r_{\rm
cut}$.

3. Each PBH is located in its host UCMH (\citealt{Mack07}). We emphasize
that the impact of X-ray emission from PBH host UCMH systems is
limited by the low abundance of PBHs ($f_{\rm PBH}\ll f_{\rm UCMH}$),
the average inefficient radiation ($\eta\ll1$), the
photon trapping effect by the accreted baryons in host UCMHs, and outflows produced by
rapid accretion feedback. Sufficient massive host UCMHs can accrete and thermalize the infalling
baryons, which are accumulated inside the UCMHs with a mass fraction of the UCMH $f_{b}>10^{-3}$,
and trap X-rays from the accreting PBHs until a critical redshift $z_{m}\sim32(\delta
m/M_{\rm PBH})^{1/2}m_{\chi,100}^{-5/12}$, below which
X-rays from a super-Eddington accretion flows onto PBHs
could escape the surrounding baryon environment in the host UCMHs. Although
the PBH abundance is $f_{\rm PBH}\ll f_{\rm UCMH}$ due to the much higher perturbation threshold
for the PBH formation, X-ray emission
could dominate over UCMH annihilation and become more promising cosmic energy source of the IGM ionization and heating at $z<z_{m}$
if the PBH abundance is above a threshold $\eta f_{\rm PBH}/f_{\rm UCMH}\sim 3.1\times10^{-8}m_{\chi,100}^{-4/3}(1+z)$,
which is only allowed beyond the standard Gaussian density perturbation scenario.

4. As UCMHs are expected to exist in our Galaxy, we expect that UCMHs collapse with the homogenous dark matter
background to form the first large scale dark matter objects (dark haloes). If this
is the case, the dark matter annihilation from UCMHs inside the first
dark halo still dominates over the extended dark matter annihilation
background inside the halo even after the halo virialization. UCMH
radiation, including both dark matter annihilation and
accretion emission, can dramatically suppress the formation of the low mass first baryonic
structure, since UCMH radiation heats the IGM and provide a hot ambient gas
environment up to $T_{\rm m}\sim10^{4}$ K.
The UCMH radiation enhances the baryon chemical quantities such as $x_{\rm ion}$ and $f_{\rm H_{2}}$ by orders of
magnitude from $x_{\rm ion}\sim10^{-6}$ and $f_{\rm H_{2}}\sim10^{-4}$
to the upper bound of $x_{\rm ion}\sim10^{-4}$ and $f_{\rm H_{2}}\sim5\times10^{-3}$.
However, the impact of UCMH radiation on the baryon temperature of the first baryonic objects
is very small, which shows that the the influence of UCMH radiation on the temperature of
first baryonic objects is small compared to the molecular hydrogen cooling and virialization time $z_{\rm vir}$.
However, the higher abundant $x_{\rm ion}$  and $f_{\rm H_{2}}$ provided by UCMH radiation
decrease the gas temperature in the later gas collapse phase and
can produce lower fragmentation mass scale and lower mass first stars.

Also, we point out that, different spectral energy distributions of
UCMH radiation also affect the processes of ionizing radiation and
heating gas. More concentrated UCMH distribution within a dark
matter halo provides a more promising ionization phenomenon of the gas in
first dark haloes. UCMHs should be distinguished from the small dark
structure which formed during the structure formation epoch after
$z\sim100$. Future work need to be done to investigate the importance of
small dark matter structure down to the earth mass compared with UCMH
radiation in the early Universe at $z\sim60$.
Also, the CMB polarization anisotropies, 21 cm spectrum and Compton
y-parameter affected by the UCMH radiation also need to be further studied
for a better constraint on the UCMH abundance.

\section*{Acknowledgments}
DZ is grateful to John Beacom and Brain Lacki for stimulating
discussions and useful comments on the manuscript.  He would also
like to thank David Weinberg and Alexander Belikov for helpful
discussions on the baryonic fraction within dark haloes and dark
halo structure formation history. Furthermore, he acknowledges Todd
Thompson and Zi-Gao Dai.

\appendix
\section{Profiles of First Dark Structure}

The dark matter profile we use in this paper are as follows. Most of
the equations list below can be found in \cite{Tegmark97},
\cite{Ripamonti07} and \cite{RMF07a}. A dark matter halo with a mass
$M_{\rm DM}$ is assumed to distribute inside a truncation radius
$R_{\rm tr}$, which is given by
\begin{equation}
R_{\rm tr}(z,z_{\rm vir})=\left\{
\begin{array}{l}\left[\frac{3}{4\pi}\frac{M_{\rm DM}}{\rho_{\rm th}(z)}\right]^{1/3}\qquad\quad\quad z\geq z_{\rm ta}\\
R_{\rm vir}\left[2-\frac{t(z)-t(z_{\rm ta})}{t(z_{\rm vir})-t(z_{\rm ta})}\right]\;\;\;z_{\rm vir}\leq z\leq z_{\rm ta}\\
R_{\rm vir}\qquad\qquad\qquad\quad\quad\;\; z\leq z_{\rm vir}
\end{array}
\right.
\end{equation}
Here $z_{\rm vir}$ is the redshift of halo virialization, $z_{\rm
ta}\simeq1.5(1+z_{\rm vir})-1$ is the turnaround redshift. The dark
matter mass inside the halo $M_{\rm DW}$ is given as a parameter
here. Dark matter within the halo is assumed to distribute uniform
at $z>z_{\rm ta}$ with a density evolution
\begin{equation}
\rho_{\rm th}(z,z_{\rm vir})=\rho_{\rm DM}(z)e^{1.9A/(1-0.75A^{2})},
\end{equation}
where $A(z)=(1+z_{\rm vir})/(1+z)$. The virial radius $R_{\rm vir}$
is given by
\begin{equation}
R_{\rm vir}=\frac{1}{2}R_{\rm tr}(z_{\rm ta})=\frac{1}{2}\left[\frac{3}{4\pi}\frac{M_{\rm DM}}{\rho_{\rm th}(z_{\rm ta})}\right]^{1/3}.
\end{equation}
For an isothermal halo profile with a core radius $R_{\rm core}$
inside $R_{\rm tr}$, the dark matter density $\rho(z)$ is a constant
$\rho_{\rm core}$ for $r\leq R_{\rm core}$, $\rho(z)\propto r^{-2}$
for $R_{\rm core}\leq r\leq R_{\rm tr}$. The density goes to the
background dark matter density for $r> R_{\rm tr}$. Here the core
radius $R_{\rm core}$ is obtained as
\begin{equation}
R_{\rm core}(z,z_{\rm vir})=\left\{
\begin{array}{l}
R_{\rm tr}(z)\qquad\qquad\qquad\qquad\qquad\;\; z\geq z_{\rm ta}\\
R_{\rm vir}\left[2-(2-\xi)\frac{t(z)-t(z_{\rm ta})}{t(z_{\rm vir})-t(z_{\rm ta})}\right]\;z_{\rm vir}\leq z\leq z_{\rm ta}\\
\xi R_{\rm vir}\qquad\qquad\qquad\qquad\qquad\;\;\; z\leq z_{\rm vir}
\end{array}
\right.
\end{equation}
where the coefficient $\xi$ is introduced as a parameter. And
$\rho_{\rm core}$ can be obtained integrating the halo mass within
$R_{\rm tr}$ as $M_{\rm DM}$. The total luminosity produced by the
annihilation of dark matter particles distributed within the
isothermal halo is
\begin{equation} L_{\rm ext,iso}=\frac{2\pi \rho_{\rm
core}^{2}}{m_{\chi}}\langle\sigma v\rangle c^{2}R_{\rm
core}^{2}\left(R_{\rm tr}-\frac{2}{3}R_{\rm core}\right).
\end{equation}

On the other hand, the widely used NFW dark matter profile is
$\rho(r)\propto r^{-1}(1+r/R_{\rm core})^{-1}$ for $r\leq R_{\rm
tr}$. Similarly we can write down the complete formula of density
distribution as the function of redshift, and the luminosity
produced by dark matter annihilation.

The virial temperature of a halo $T_{\rm vir}$ is calculated as
\begin{equation}
T_{\rm vir}=\frac{\mu m_{p}}{2k_{B}}\frac{G M_{\rm DM}}{R_{\rm vir}},
\end{equation}
sometimes people use the total mass $M_{\rm halo}$ instead $M_{\rm
DM}f_{\chi}$. However, as we consider if the gas from the background
can collapse into the halo, we adopt the pure dark halo mass as to
calculate the initial virial temperature. Therefore we can obtain
the virial temperature as $T_{\rm vir}=8.2\times10^{-3}$ K
$(1+z_{\rm vir})(M_{\rm DM}/M_{\odot})^{2/3}$, or $T_{\rm vir}=380$
K $(M_{\rm DM}/10^{4}M_{\odot})^{2/3}(1+z_{\rm vir})/100$. The
baryon falling can occur when the IGM temperature $T_{\rm m}$ around
the halo to be $T_{\rm m}<T_{\rm vir}$, otherwise the gas pressure
will impede the gas falling into the halo and collapsing to form
smaller structure. Moreover, the condition $T_{\rm CMB}=T_{\rm vir}$
with $T_{\rm CMB}=2.73$ K $(1+z_{\rm vir})$ gives a critical halo
mass $M_{\rm DM}=6.1\times10^{3}M_{\odot}$.

\end{document}